\documentclass[11pt]{article}

\pdfoutput=1
\usepackage{color}
\usepackage{epsfig, palatino}
\usepackage{pstricks,pst-node,pst-tree}
\usepackage{epic}
\usepackage{mathrsfs}
\usepackage{ae} 
\usepackage[T1]{fontenc}
\usepackage[ansinew]{inputenc}
\usepackage{amsmath}
\usepackage{amssymb}
\usepackage{graphicx}
\usepackage{ulem}
\usepackage{simpler-wick}
\definecolor{darkblue}{cmyk}{0.9,0.9,0,0}
\usepackage[colorlinks=true,linkcolor=darkblue,citecolor=darkblue,urlcolor=darkblue]{hyperref}
\usepackage{cite}
\usepackage{hyperref}
\usepackage{wasysym}
\usepackage{varioref}
\usepackage{makeidx}
\usepackage[english]{babel}
\usepackage{simplewick}
\usepackage{array}
\usepackage{multirow}
\usepackage{tikz}
\usetikzlibrary{arrows.meta}
\usepackage{cancel}
\usepackage{subcaption}

\newcounter{mycounter} 

\usepackage{lipsum}

\usepackage{animate}
\usepackage{mdframed}
\usepackage[font={small}]{caption}
\usepackage{esint}

\newcommand{\comment}[1]{}

\newcommand{\beq}{\begin{equation}}
\newcommand{\eeq}{\end{equation}}
\newcommand{\beqq}{\begin{equation*}}
\newcommand{\eeqq}{\end{equation*}}
\newcommand\beqa{\begin{eqnarray}}
\newcommand\eeqa{\end{eqnarray}}
\newcommand\beqaa{\begin{eqnarray*}}
\newcommand\eeqaa{\end{eqnarray*}}
\newcommand\bea{\begin{array}}
\newcommand\eea{\end{array}}

\newcommand{\nn}{\nonumber}

\newcommand{\diag}[1]{{\rm diag}(#1)} 
\newcommand{\neqa}{\nonumber\end{eqnarray}} 
\newcommand{\la}[1]{\label{#1}}

\renewcommand{\d}{\partial}

\newcommand{\<}{{\langle}}
\renewcommand{\>}{{\rangle}}

\newcommand{\re}{\relax{\rm I\kern-.18em R}}

\renewcommand{\sp}{p\hspace{-.40em}/}

\definecolor{darkgreen}{rgb}{0.0, 0.45, 0.0}

\def\tr{{\rm\,tr\,}}
\def\Tr{{\rm\,Tr\,}}

\def\su2{{SU(2)}}

\def\a{{\alpha}}

\def\[{\left[}
\def\]{\right]}

\def\s{\sigma}
\def\a{\alpha}

\def\({\left(}
\def\){\right)}
\def\[{\left[}
\def\]{\right]}

\def\<{\langle}
\def\>{\rangle}

\def\i2{\frac{i}{2}}

\def\spi{\relax{\rm \pi\kern-0.5em /}}
\def\sA{\relax{\rm A\kern-0.5em /}}
\def\sp{\relax{\rm p\kern-0.5em /}}
\def\sd{\relax{\rm \d\kern-0.5em /}}
\def\sk{\relax{\rm k\kern-0.5em /}}
\def\sn{\relax{\rm n\kern-0.5em /}}
\def\sl{\relax{\rm l\kern-0.5em /}}
\def\sP{\relax{\rm P\kern-0.7em /}}
\def\sBethe{\relax{\rm \Bethe\kern-0.5em /}}

\def\2F1{\,_2{\rm F}_1}

\makeindex

\begin{document}

\thispagestyle{empty}

\renewcommand{\thefootnote}{\fnsymbol{footnote}}
\setcounter{page}{1}
\setcounter{footnote}{0}
\setcounter{figure}{0}

\begin{center}
$$$$
{\Large\textbf{\mathversion{bold}
Universality of Heavy Operators in Matrix Models
}\par}

\vspace{1.0cm}

\vspace{1.0cm}

\textrm{Andrea Guerrieri$^\text{\tiny 1}$\footnote{\tt  andrea.leonardo.guerrieri@gmail.com}, Harish Murali$^\text{\tiny 2,\tiny3}$\footnote{\tt  harish02murali@gmail.com}, Pedro Vieira$^\text{\tiny 2,\tiny 4}$\footnote{\tt  pedrogvieira@gmail.com}}

\vspace{1.2cm}
\footnotesize{\textit{
$^\text{\tiny 1}$ Department of Mathematics, City St. George's, University of London\\ Northampton Square, EC1V 0HB, London, UK
\\
$^\text{\tiny 2}$Perimeter Institute for Theoretical Physics,
Waterloo, Ontario N2L 2Y5, Canada\\
$^\text{\tiny 3}$Department of Physics and Astronomy, University of Waterloo, Waterloo, Ontario, N2L 3G1, Canada\\
$^\text{\tiny 4}$ICTP South American Institute for Fundamental Research, IFT-UNESP, S\~ao Paulo, SP Brazil 01440-070
}  
\vspace{4mm}
}

\par\vspace{1.5cm}

\textbf{Abstract}\vspace{2mm}
\end{center}

\noindent \\
In large $N$ theories with a gravity dual, generic heavy operators should be dual to black holes in the bulk. The microscopic details of such operators should then be irrelevant in the low energy theory. We look for such universality in the strong coupling limit of a very simple two matrix model -- the Hoppe model. Using analytics as well as Monte Carlo simulations, we show that there exists a universal black hole regime where the eigenvalue densities are given by parabolas and the correlation functions of probes in these backgrounds are completely determined by a few parameters. An important feature of strong coupling in this model is that the matrices commute and one can define joint eigenvalue distributions which also exhibit universality. These two results extend the beautiful findings of Berenstein, Hanada and Hartnoll  \cite{Berenstein:2008eg}.
Not all heavy operators are universal and at strong coupling there is a sharp phase boundary between the universal and non universal regimes (Of course this should not be confused with the universality of eigenvalue spacing in matrix models). Moreover, in the non universal phase, we also find an interesting phenomenon we call Abelianization where some eigenvalues run off to infinity, reminiscent of heavy dual giant gravitons in $\mathcal N=4$ SYM.



% ...




 \newpage

 \tableofcontents


 \newpage
\setcounter{footnote}{0}
\renewcommand\thefootnote{\arabic{footnote}}
\renewcommand\thempfootnote{\arabic{mpfootnote}}
\section{Introduction}\label{introSection}
The precise form of a quantum microstate of a macroscopic piece of coal, or the precise form of a quantum microstate of a macroscopic black-hole is irrelevant. Many such microstates ought to give the very same macroscopic physics. Colloquially, we can call it universality. 

This paper is about looking for similar universality in some of the simplest possible physical toy models. It is important that these toy models not be too simple. Solvable models, almost by definition, are non-chaotic and thus often miss the simple universality we are after. An example would be one matrix models or integrable models. We suggest that some multi matrix models at strong coupling might be good toy models for such universality.\footnote{Note that some of the most exotic examples of holography such as the polarized IKKT model relate quantum gravity to matrix models of the sort studied in this paper (albeit with many more bosonic as well as fermionic matrices!). Ultimately, we hope to transport some of the understanding of this paper to those models in the future.} Note that the universality we are after is very different from the universality of eigenvalue spacings and critical edge distributions typically seen in 2d gravity, see e.g. \cite{Brezin:1990rb,Mehta2004,Eynard:2015aea}.

For this work, our favorite observable will be the evaluation of probes in the so-called Hoppe model \cite{hoppe:1989} at large number of colours $N\to \infty $ and strong coupling $\lambda \to \infty $ in the presence of a huge operator, 
\beq
\< O_\text{probe}  \> = \frac{\int dX dY O_\text{probe}(X,Y) O_\text{huge}(X,Y)  \exp\!\Big(\!-N\, \text{tr}\big(\displaystyle \tfrac{1}{2} X^2+\tfrac{1}{2}Y^2 - \lambda [X,Y]^2\big)\Big)}{\texttt{same thing with $O_{\text{probe}}(X,Y)\to 1$} } \,. \label{mainEq1}
\eeq
The operator $O_\text{huge}(X,Y)$ is an operator which is big enough to deform the vacuum distribution of the matrices $X$ and $Y$ in the large $N$ limit. For example, we could take $O_\text{huge}(X,Y)$ to be a classical source in one of the two directions, say $O_\text{huge}(X,Y)=e^{N\, \tr\!(JX)}$ where $J$ is some diagonal matrix with real entries $J_i$. 

The operator~$O_\text{probe}(X,Y)$, on the contrary, is an operator which does not scale with $N$ as strongly and therefore does not backreact on the geometry. For example we could have~$O_\text{probe}(X,Y)=\text{tr}(X^n)$ so that in this example the probe would measure the moments of the eigenvalue distribution~$\rho(x)$ of the matrix $X$. 

Lets carry out some illustrative computation in this particular example where both the source and the probe operators just depend on one of the two matrices~($X$). In this case the other matrix~($Y$) can be analytically integrated out since it appears in a gaussian way and the remaining integral can be nicely reduced to an integral over the $X$ eigenvalues $x$ to yield
\beq
\left\<{ \color{blue}\frac{1}{N} \text{tr}(X^n)} \right\> = \frac{\int dx_1\dots dx_N \, {\color{blue}x_1^n}\, \Delta_\lambda(x)^2 \exp\Big(-\tfrac{N}{2} \sum_{i=1}^{N} x_i^2\Big) {\color{red}\big(\det_{i,j} e^{N J_i x_j})/\Delta(x)}
}{\texttt{same thing with ${\color{blue}x_1^n \to 1}$} } \,. \label{mainEq}
\eeq
The red term comes from integrating out the angular part $U$ of $X=UxU^\dagger$ which shows up in the classical source term only. The integral over $U$ can be explicitly done; it is the Harish-Chandra-Itzikson-Zuber (HCIZ) integral and leads to the red term  \cite{HC,IZ}. The integration over the $Y$ matrix modifies the usual matrix model Vandermonde factor $\Delta(x) \equiv \prod_{i<j}(x_i-x_j)$ into the only factor where the coupling $\lambda$ appears, namely
\beq
\Delta_\lambda(x)^2 = \prod_{i< j} \frac{(x_i-x_j)^2}{1+2 \lambda (x_i-x_j)^2} \nn
\eeq
Because of the huge operator the density of $x$ eigenvalues $\rho_{J}(x)$ gets deformed from the vacuum case $\rho_0(x)$. The probes will then probe the moments of this new density, 
\beq
\left\<{ \color{blue}\frac{1}{N} \text{tr}(X^n)} \right\> = \int \rho_J(x) x^n \,. \nn
\eeq
We can now go back to the motivation of the opening paragraph and ask ourselves: \textit{Is there any universality here?} In short, what we find is:

{At weak or finite coupling $\lambda$ there is \textit{no} universality: different big operators (i.e. different sources $J$ in this example) lead to different densities and different probe moments. No surprise here. However, at strong coupling there are often two phases. In one phase there \textit{is} universality: different sources yield the same density $\rho_J$ (up to some trivial rescalings) and thus to the very same moments (up to some trivial rescalings). In the other phase this universality does not seem to be there.} 
\begin{figure}
\begin{centering}
\includegraphics[width=0.85\linewidth]{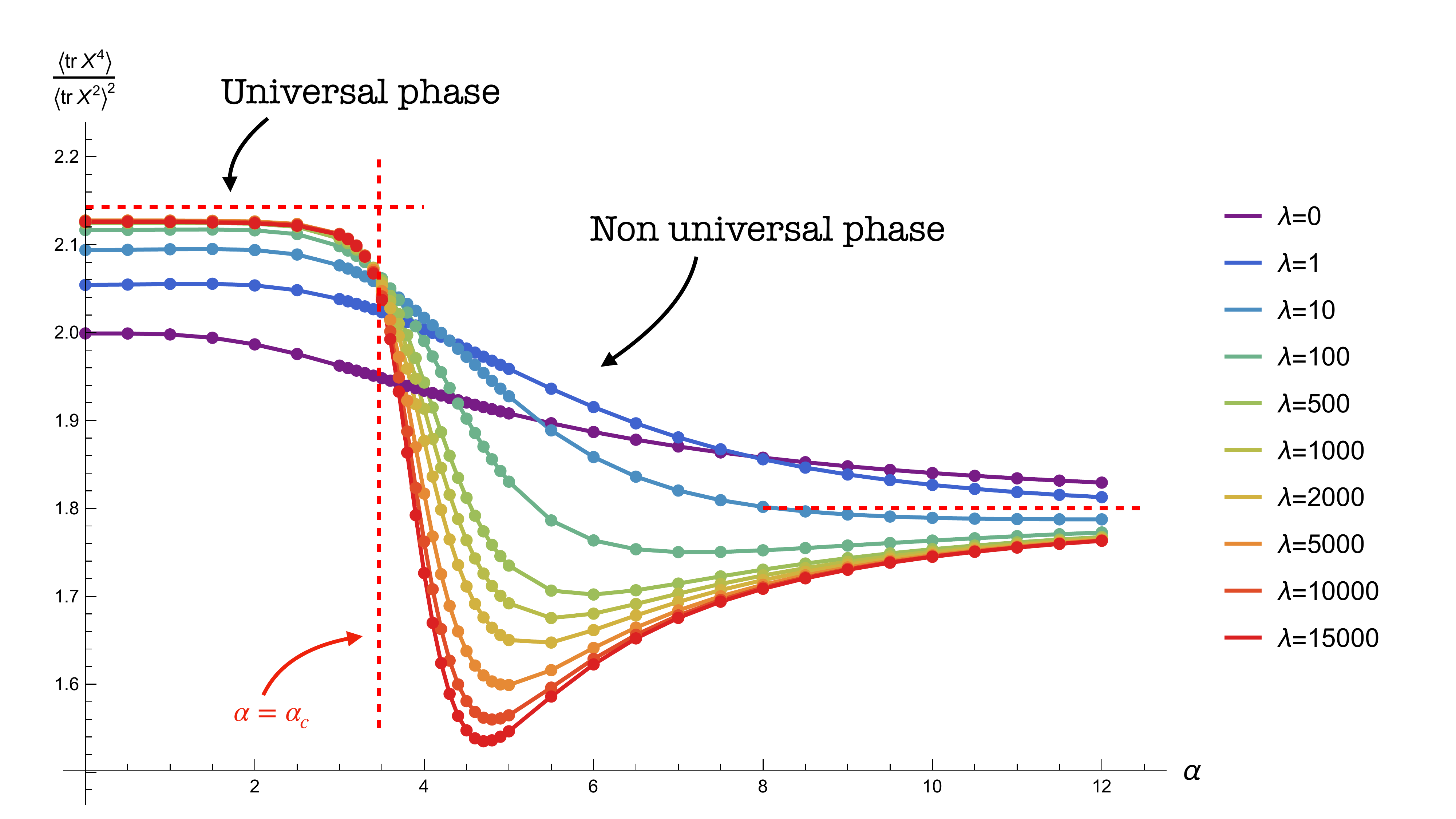}
    \caption{A plot of moments for uniform line source $J$ of width $\alpha$. For $\alpha<\alpha_c$ all curves approach the same universal plateau. Highlighting this universality is the main purpose of this work; it shows up at strong coupling only. For $\alpha>\alpha_c$ this universality is gone. At large $\alpha$ a different, simpler, sort of universality emerges. That simpler universality of large sources is valid at any coupling.}
    \label{panel1}
    \end{centering}
\end{figure}
Figure \ref{panel1} illustrates this in the simple example where the classical source $J$ consists of $N$ eigenvalues $J_j$ equally spaced from $-\alpha/2$ to $\alpha/2$. At weak coupling clearly all densities depend non-trivially on the value of $\alpha$. (We compute these weak coupling density moments analytically in the next section.) At strong coupling for 
\beq
\alpha \le \alpha_c= \sqrt{12} \nn
\eeq
we see that all densities are the same up to a trivial rescaling! We call this phase the universal phase. ($\diamond$ We will sometimes call it the \textit{Black Hole} phase. $\diamond$ \footnote{Of course, this is colloquial overreach at this point since the Hoppe model is very much a toy model. Still, it is too tempting to avoid this and other similar gravitational analogies. We will try to bracket such poetic liberties with two grain of salt symbols $\diamond$ which the more conservative readers can safely jump.}) 
For $\alpha>\sqrt{12}$ the density is no longer universal. We call this the non-universal phase. We derive the location of this strong coupling phase transition -- and generalize it to a generic source $J$ -- in the next sections. 

In other words, we find a remarkable simplicity emerging at strong coupling. We believe that this emergent simplicity stems from the important fact that at strong coupling the matrices $X$ and $Y$ commute with each other. This commutativity is a well known but non-trivial fact which we review below. Were $N$ finite, this would effectively mean we could diagonalize both matrices at once. But $N$ is infinite and the business of commuting infinite matrices is subtle as we discuss below. 

This preliminary exploration begs the question of how general is this universality. What happens if the huge operator and/or the probe operator depends on both matrices $X$ and $Y$; can we still find some universality there? Is there some similar universality for other classes of huge operators? Can we find other multi-matrix models with similar commutativity and universality at strong coupling? And are there special huge operators for which another sort of universality arises? The answer to all these questions is \textit{yes} and we turn to them sequentially in the next sections.

\section{Hoppe Model with Homogeneous Source: Very Instructive Example}\label{uniformLineSection}
In the previous introductory section we presented figure \ref{panel1} without any details. In this section we will provide several details on how to analytically understand many of the features of this plot. This constant classical source example is highly suggestive of the most general picture which is why we want to cover it in excruciating detail. The reader interested in the more general case of other more generic huge operators with more general sources -- or even huge operators of a totally different form -- can skip to the next section in a first reading. 

To remove some clutter, in what follows we will sometimes introduce a normalized trace $\Tr(\dots)$ related to the usual trace~$\tr(\dots)$ as 
\beq
\Tr(\dots)=\frac{1}{N}\tr(\dots) \,. \nn
\eeq

\subsection{Weak Coupling}

We start with the purple curve in figure \ref{panel1} corresponding to the zero coupling limit $\lambda=0$. When $\lambda=0$ the model is gaussian. It remains gaussian even in the presence of classical sources given by a huge operator of the form $O_\text{huge}=e^{N\,\text{tr}(J_X X+J_Y Y)}$. We can therefore exactly integrate the matrices $X$ and $Y$ in the presence of such sources and compute moments of the matrices as usual, by taking derivatives of the result with respect to these sources, 
\beq
\left\< \tr\!({\color{blue}XX}{\color{red}Y}{\color{blue}X}{\color{red}Y}\dots) \right\>= e^{-\frac N2 \text{tr}(J_X^2+J_Y^2)}\ \tr\Big( {\color{blue}\frac{d}{dJ_X} \frac{d}{dJ_X}}{\color{red} \frac{d}{dJ_Y}}{\color{blue} \frac{d}{dJ_X} }{\color{red}\frac{d}{dJ_Y}} \dots \Big) e^{\frac{N}{2} \text{tr} (J_X^2+J_Y^2)} \, \nn
\eeq
where the derivatives only act to the right. In particular, 
\beq
\frac{\left\< \Tr X^4\right\>}{\left\< \Tr X^2\right\>^2} = \frac{\Tr J_X^4+4\Tr J_X^2+2(\!\Tr J_X)^2+2}{\left(1+\Tr J_X^2\right)^2}  \nn
\eeq
which evaluates to 
\beq
\frac{\left\< \Tr X^4\right\>}{\left\< \Tr X^2\right\>^2} =\frac{2+\alpha^2/3+\alpha^4/80}{(1+\alpha^2/12)^2} \label{alphaLambdaZero}
\eeq 
for a classical source $J_X$ with eigenvalues uniformly spaces between $-\alpha/2$ and $\alpha/2$. This perfectly matches with the purple curve in figure \ref{panel1}. 

\begin{figure}[t]
\begin{centering}
\includegraphics[width=0.8\linewidth]{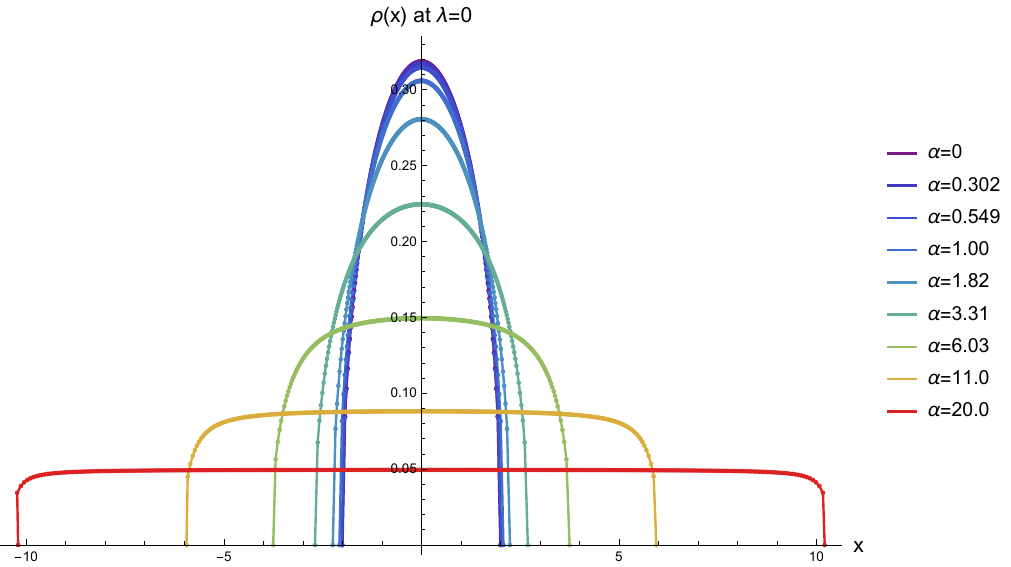}
    \caption{At zero coupling, $\lambda=0$, the densities interpolate smoothly between the semi-circle rule at vanishing source to a flat constant distribution for large source. (Here $N=300$ and the last point for each curve - where the density vanishes - was estimated through a fit.)}
    \label{densitiesZeroCoupling}
    \end{centering}
\end{figure}

We should stress two trivial but important things. First although the computation of these moments is straightforward, the corresponding density of eigenvalues $\rho(x)$ which dominate the saddle point in (\ref{mainEq}) is very non-trivial. For the equally spaced eigenvalues case where $J_k=-\alpha/2+k \alpha/N$, the red term in (\ref{mainEq}) simplifies greatly into a ratio of Vandermondes, 
\beq
\det_{i,j}\,\frac{e^{N J_i x_j}}{\Delta(x)} = e^{-\frac {N\alpha}2 \tr X}\ \frac{\Delta\left(e^{\alpha x}\right)}{\Delta(x)}\label{homoLineVandermondeRatio}
\eeq
and therefore the saddle point equations arising from (\ref{mainEq}) would in this case simply read 
\beq
-x-\frac\alpha2 + 2\fint dy\, \frac{\rho(y)}{(x-y)(1+2\lambda(x-y)^2)} + \int \rho(y)\left(\frac{\alpha e^{\alpha x}}{e^{\alpha x}-e^{\alpha y}}-\frac1{x-y}\right) =0 \label{SPalpha}
\eeq
Note that here we kept $\lambda \neq 0$ for future reference. When $\lambda=0$ this equation remains very non-trivial. As $\alpha$ interpolates between $0$ and $\infty$ the density interpolates from the famous Wigner semi-circle at zero sources to a constant density at very large sources, see figure \ref{densitiesZeroCoupling}. As a check note that 
\beq
\frac{\displaystyle\int\limits_{-2}^2 dx\,\frac{\sqrt{4-x^2}}{2\pi}\,x^4 }{\Big(\texttt{same with } x^4 \to x^2 \Big)^{ 2}} = 2  \,, \qquad \frac{\displaystyle\int\limits_{-\alpha/2}^{\alpha/2} dx\,\frac{1}{\alpha} x^4}{\Big(\texttt{same with } x^4 \to x^2 \Big)^{ 2}} = \frac{9}{5} \,,  \nn
\eeq
precisely matching (\ref{alphaLambdaZero}) at $\alpha=0$ and $\alpha=\infty$ respectively. 
For finite $\alpha$ the resolvent $G(x)=\int \rho(y)/(x-y)$ will have a cut in the physical sheet which grows as $\alpha$ grows and an infinite ladder of cuts separated by $2\pi i/\alpha$ in the second sheet. It would be interesting to work it out analytically but we will not need it here. For us it suffices to point out that this is clearly a very rich object with a rich $\alpha$ dependence and thus no universality. Indeed, the purple curve in figure \ref{panel1} given by~(\ref{alphaLambdaZero}) depends non-trivially on $\alpha$ for any range of $\alpha$.

\subsection{Strong Coupling}

This is in stark contrast with what we see in that same figure at strong coupling! As $\lambda$ grows we see that there is a finite universal region for $\alpha$ below some critical value where the moments become universal and independent of $\alpha$, signaling the emergence of universality in this regime. We now turn to the strong coupling analysis. 

We start with the vacuum, that is $J_X=J_Y=0$. The vacuum was studied at any value of the coupling by Hoppe in \cite{hoppe:1989} and later by Kazakov, Kostov and Nekrasov \cite{Kazakov:1998ji}. The solution is quite rich involving elliptic functions. At strong coupling the result simplifies greatly as pointed out by Berenstein, Hanada and Hartnoll \cite{Berenstein:2008eg}. The key is to observe that for a smooth density we have
\beq
\fint \frac{\rho(y)}{(x-y)(1+2 \lambda (x-y)^2)} = -\frac{\pi}{\sqrt{2\lambda}} \rho'(x)+\ldots\label{strongCouplingDensity}
\eeq
so that the saddle point equation in the absence of sources simply becomes 
\beq
-x-\frac{2\pi}{\sqrt{2\lambda}}\rho'(x)=0\label{vacuumHoppeSPE}
\eeq
so that the leading order density is a simple parabola \cite{Berenstein:2008eg}
\beq
\rho(x)=\frac{L^2-x^2}{4L^3/3} \,, \label{parabola}
\eeq
with 
\beq
L=L_\text{vacuum} \equiv \frac{(3\pi)^{\frac13}}{(2\lambda)^{\frac16}} \label{Lvac}
\eeq
This density can be corrected order by order following \cite{Filev:2013pza} see also appendix A in \cite{Berenstein:2008eg}. Importantly, note that $L$ is very small so this is a very peaked density. Let us do a quick check of this result. For a parabola, we have
\beq
\frac{\int_{-L}^L \frac{L^2-x^2}{4L^3/3} x^4}{\big(\int_{-L}^L \frac{L^2-x^2}{4L^3/3} x^2\big)^2} = \frac{15}{7} \simeq 2.14 \label{ratioMomentaParabola}
\eeq
which is precisely the value of the first dashed curve in the figure which the curves neatly approach at small $\alpha$, and in particular for $\alpha=0$.\footnote{
The numerical curves do not lie on top of this dashed curve because of the finite coupling corrections. In this model reaching strong coupling numerically is tough because of the funny $1/6$ powers to which the coupling is raised, see \eqref{Lvac}. For a huge $\lambda=15000$ we get $L_\text{vac} \simeq 0.38$ which is of course not that tiny. And the issue is that we can not simply add a few more zeroes of $\lambda$ because we need to take strong coupling while staying in the large $N$ limit! In other words, we take $N\to \infty$ first and $\lambda\to \infty$ second. To generate figure \ref{panel1} we needed to use $N=1000$ and solve a thousand coupled saddle point equations for the location of the eigenvalues $x_j$. To increase $\lambda$ further while staying well inside the large $N$ limit we would also need to increase $N$ further which becomes too expensive.}  

\begin{figure}[t]
    \centering
    \includegraphics[width=\linewidth]{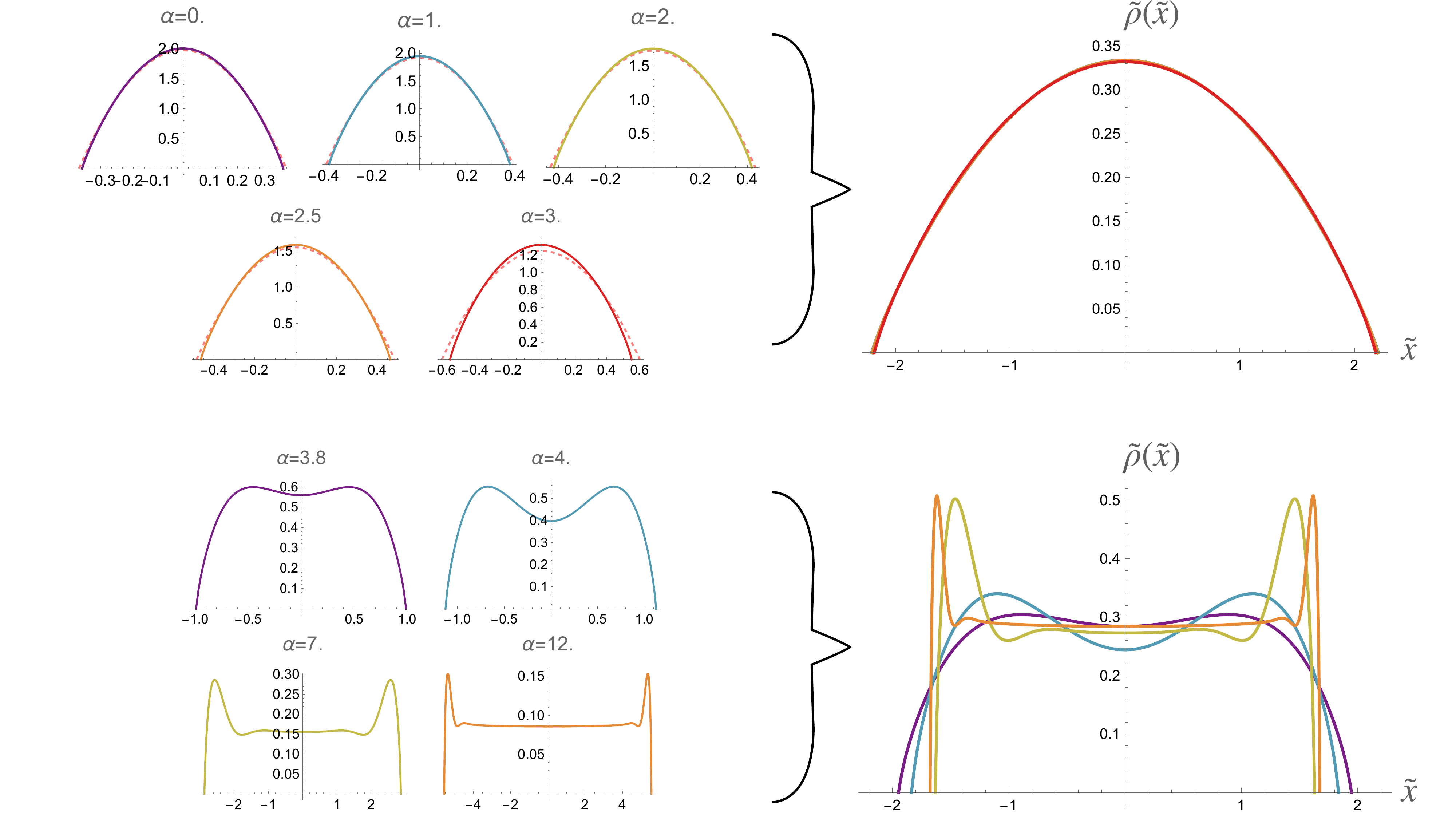}
    \caption{Density $\rho(x)$ for the line source for various $\alpha$'s at strong coupling $\lambda=15000$. The densities are computed by solving the discretized version of \eqref{SPalpha} with $N=1000$ points -- these are the solid lines on the left. On the right, we rescale the points to $\tilde x={x}/{\sqrt{\tfrac{1}{N}\langle\tr X^2\rangle}}$ such that they have unit variance. In the top row, we are in the universal regime $\alpha\leq \alpha_c\approx3.46$ and after rescaling, the densities collapse to the same curve! The densities also agree well with analytic parabolas \eqref{LengthAlpha} shown as dashed red lines. Close to criticality, the agreement is not perfect due to finite coupling effects. In the bottom row, we are in the non-universal regime and the support of the densities is no longer $O(\lambda^{-\frac16})$. We see here that the distribution is approaching~$\rho(x)=1/\alpha$ at large $\alpha$.}
    \label{densitiesStrong}
\end{figure}

Now, as we see in the figure, the strong coupling curves form a plateau and approach the dashed line not only for $\alpha=0$ but actually for a finite range of $\alpha$. This hints at some sort of ``parabola universality" in this regime and indeed that is precisely what we have. To see this let us turn on $\alpha$ and assume that the density is still a parabola given by (\ref{parabola}) with a yet to fix $L(\alpha)$. This assumption is indeed self-consistent. Indeed, a short computation shows that the integral
\begin{multline}
\fint_{-L}^L dy \left(\frac{2}{(x-y)(1+2 \lambda (x-y)^2)}+\frac{\alpha  e^{\alpha  x}}{e^{\alpha  x}-e^{\alpha  y}}-\frac{1}{x-y}
\right) \frac{L^2-y^2}{4L^3/3}  =\\ \frac\alpha2 + \left(\frac{\alpha^2}{12} + \frac{L_{\text{vacuum}}^3}{L^3}\right) x + O\left(\frac1{\sqrt\lambda}\right)\nn
\end{multline}
so that (\ref{SPalpha}) is perfectly solved as long as we set 
\beq
L(\alpha)=L_\text{vacuum} \times \frac{1}{\(1-\frac{\alpha^2}{12}\)^{1/3}} \label{LengthAlpha}
\eeq
We see that the density is still a nice parabola albeit with a renormalized length. That is why the normalized ratio of moments perfectly matches the $L$ independent parabola value~$15/7$  computed above (\ref{ratioMomentaParabola}). It is in this sense that these heavy operators sourced by $J_X$ appear universal at strong coupling. At the same time, the renormalized length expression (\ref{LengthAlpha}) also signals the breakdown of the validity of the parabola approximation. It clearly fails at $\alpha=\alpha_c=\sqrt{12}$ when the parabola length blows up. Above this value the density can no longer be a parabola. This is indeed precisely what we observe numerically, see figure \ref{densitiesStrong}.\footnote{The reader might be thinking. ``Oh, indeed, for $\alpha>\alpha_c$ the density can not be a parabola. What is it then? It should still be simple and computable at strong coupling right?". We agree dear reader. Tragically, we failed to find it analytically. } 

In sum: we have two regimes. For $\alpha<\alpha_c$ the density is universal. It is a sharply peaked parabola parametrized by a single number $L$ which scales as $\lambda^{-1/6}$. Motivated by our goal to find thermal-like universality in matrix models, we could dub this universal phase the \textit{black-hole phase}. In the same way that a black-hole is parametrized by a tiny number of parameters -- due to no hair theorems -- so are the eigenvalue densities below $\alpha_c$.
For $\alpha>\alpha_c$ the density is no longer universal as clearly seen in the figure. Were we brave enough and we would perhaps dub this phase as the \textit{star} phase and draw an analogy between the continuously changing non-universal eigenvalue densities in this regime and the existence of more exotic solutions (as opposed to BHs) with several continuous moduli. 

\subsection{Large Sources} \label{largeSourcesSec}

Let us quickly point out to another very different kind of universality which we observe for very large $\alpha$: all curves in figure \ref{panel1} clearly approach the same asymptote for large sources. This universality is \text{not} related to the ``black hole" universality just discussed above which is the main topic of this work. Suppose we insert an operator of the form $O_\text{huge}(X,Y)=e^{N\,\text{tr}(J_X X+J_Y Y)}$ in the path integral with very large sources $J_X$ and $J_Y$. If the sources are huge we expect the dynamics of the matrices to be dominated by solutions to the equations of motion 
\beq
X-J_X=\lambda Y\[X,Y\] \,, \qquad Y-J_Y=\lambda X\[Y,X\] \,. \nn
\eeq
These equations have many solutions. If $J_X$ and $J_Y$ commute, one clear solution is simply $X=J_X$ and $Y=J_Y$. Within Hermitian saddles, this is the dominant solution since the action is a sum of traces of squares of hermitian matrices, 
\beq
S=\tfrac{1}{2}\tr(X-J_X)^2+\tfrac{1}{2}\tr(Y-J_Y)^2+ \lambda \tr(i [X,Y])^2-\tfrac{1}{2}(\tr J_Y^2+\tr J_X^2) \ge -\tfrac{1}{2}\tr J_Y^2-\tfrac{1}{2}\tr J_X^2 \nonumber
\eeq
with equality satisfied only for the solution~$X=J_X$ and~$Y=J_Y$. Therefore,  densities of eigenvalues of $x$ and $y$ should approach the densities of eigenvalues of~$J_X$ and~$J_Y$. For a homogenous big source~$J_X$ we find that $\rho(x)$ should approach a constant density between $-\alpha/2$ and $\alpha/2$ no matter the coupling. This is precisely what we observe numerically and it is the reason why all curves eventually converge to the same asymptote in figure \ref{panel1}. 

Let us stress that this universal simple behavior holds whenever we have big sources~$J_X$ and~$J_Y$ which commute with each other. It would be interesting to explore the non-commuting case as well. 

This concludes the analysis of the various notable features of figure \ref{panel1}. Before closing this section we turn to one last important property of this model which emerges at strong coupling: The matrices~$X$ and~$Y$ commute. In particular this allows us to definite meaningful joint eigenvalues distributions~$\rho(x,y)$ and uplift the parabola universality for~$\rho(x)$ observed above to this richer quantity. 

\subsection{Joint Distributions. From Parabolas to Hemispheres.} \label{parHem}

We integrated $Y$ out easily because $Y$ appears in a Gaussian way and the huge operator only depended on the matrix $X$. Without huge operator we could of course integrate out $X$ instead to conclude that the density of $Y$ is equal to the density of $X$. In the presence of the huge operator we can not simply integrate out $X$ so we follow a different route. We simply note that moments of~$Y$ can still be easily computed since the integration of $Y$ in the presence of a few $Y$'s downstairs simply adds a few Wick contractions with a simple $Y$ propagator. For example, we would get
\beq
\left\< \text{Tr}({ \color{blue}X^{n-k} {\color{red} Y} X^{k}  {\color{red} Y}}) \right\> =\frac{1}{\mathcal Z} \int dx_1\dots dx_N \, {\color{red}\sum\limits_{i,j} \frac{\color{blue} x_i^{n-k} x_j^{k}}{1+2\lambda(x_i-x_j)^2}}\, \Delta_\lambda(x)^2 \exp\Big(-\tfrac{N}{2} \sum_{i=1}^{N} x_i^2\Big) O_\text{huge}(x) 
\label{mainEqY2}
\eeq
All we used was that the huge operator only depends on $X$. In the large $N$ limit, we get 
\beq
\left\<\text{Tr}({ \color{blue}X^{n-k} {\color{red} Y} X^{k}  {\color{red} Y}}) \right\> = \int dx \int dx' \frac{x^{n-k} x'^k \rho(x) \rho(x')}{1+2\lambda(x-x')^2} \,. \label{int1}
\eeq
So far the coupling could be anything. At strong coupling the integration is localized at $x=x'$ and 
\beq
\left\<\text{Tr}({ \color{blue}X^{n-k} {\color{red} Y} X^{k}  {\color{red} Y}}) \right\> \simeq \frac{\pi}{\sqrt{2\lambda}} \int dx \, x^{n}  \rho(x)^2 \,. \label{int2}
\eeq
Note that the strong coupling localization  of the integral has an important consequence: the right hand side does not depend on $k$; only $n$ shows up! In other words, it does not matter where we put the $Y$'s in the sea of $X$'s: The matrices effectively commute! We could repeat this analysis with more $Y$'s; we would need to do more Wick contractions and see which of those Wick contractions dominate at large $N$, see appendix \ref{WickAppendix} for details. For finite coupling we end up with several integrals but at strong coupling they all collapse just as when going from (\ref{int1}) to (\ref{int2}). At the end of the day we conclude that $\left\<\text{Tr}({ \color{blue}X^{n_1} {\color{red} Y^{m_1}} X^{n_2}  {\color{red} Y^{m_2}} }\dots) \right\>$ is simply equal to
\beq
\left\<\text{Tr}({ \color{blue}X^{n} {\color{red} Y^m}}  ) \right\>  = C_{m/2} \(\frac{\pi}{\sqrt{2\lambda}}\)^{m/2} \int dx\ x^n \rho(x)^{1+m/2}  \label{sInt}
\eeq
where again only the total number of fields $n=\sum n_j$ and $m=\sum m_j$ matter and not their particular order. Here $C_k$ is the $k^{\text{th}}$ Catalan number which counts number of planar contractions of $2k$ Gaussian matrices and $m/2$ is the number of loops. (Odd $m$ moments vanish by parity.) 

Equivalently, we conclude that such moments can be computed from a joint eigenvalue distribution $\rho(x,y)$ at strong coupling as 
\beq
\left\<\text{Tr}({ \color{blue}X^{n} {\color{red} Y^m}}  ) \right\>  =  \int dx \int dy \,\rho(x,y) \, x^n y^m \label{dInt}
\eeq
where 
\beq
    \rho(x,y) = \frac2{\pi} \(\frac\lambda{8\pi^2}\)^{\frac14} \sqrt{\rho(x)-\(\frac\lambda{8\pi^2}\)^{\frac12}y^2} \,.\label{2drhoxy}
\eeq
Indeed, performing the $y$ integration in (\ref{dInt}) precisely leads to (\ref{sInt}). This holds for any density $\rho(x)$. In the homogenous source example studied above we had $\rho(x)$ given by a sharply peaked parabola for~$\alpha<\alpha_c$ or by a large delocalized non-universal distribution for~$\alpha>\alpha_c$ and this formula would hold in both regimes. After all, all we used is that the operator $O_\text{huge}(X)$ only depends on $X$ and that some smooth density $\rho(x)$ exists. 

When $\rho(x)$ is a parabola of length $L_X$, we get
\beq
    \rho(x,y) = \frac{3}{2\pi L_X L_Y}  \sqrt{1-\frac{x^2}{L_X^2} - \frac{y^2}{L_Y^2}} \,, \la{XYJointEllipse}
\eeq
where 
\beq
L_Y=\sqrt{\frac{3\pi}{L_X}}\frac1{(2\lambda)^{\frac14}} \,. \label{LyLx}
\eeq
This tells us that the distribution of $Y$ is also a parabola but with length $L_Y$. This can also be nicely backed up by numerics, see figure 
\ref{twoCenterFig}.

Note that we have radial symmetry only when $L_X=L_Y$. Solving for $L_X$ we see that this happens when $L_X=L_\text{vacuum}={(3\pi)^{\frac13}}/{(2\lambda)^{\frac16}}$, which is true for the vacuum. It makes sense: since our huge operator only depends on one of the matrices $X$, the only way to restore rotational symmetry should be to set it to the identity operator. For the vacuum, this hemisphere distribution was previously derived in \cite{Berenstein:2008eg, OConnor:2012vwc}. In the next section we will see that universal ellipsoid distributions of the form (\ref{XYJointEllipse}) abound in the presence of very generic huge operators; the lengths of the bigger and smaller semi-axes of the ellipse will, however generically no longer follow (\ref{LyLx}) once the huge operators will be more generic and depend on both matrices $X$ and $Y$. 

Having worked out this instructive example in detail, we recast our introduction closing words. Can we still find this same sort of black hole universality for much more general operators? Is the black hole phase -- with its associated parabola and hemisphere densities -- universal for those operators are well? Can the location of the phase transition between the black hole universal phase and the non-universal phase be quantitatively understood for much more general operators. For even more involved huge operators, should we perhaps employ numerical methods such as hybrid Monte-Carlo to unveil this universality? The answer to all these questions is \textit{yes} as we explain in the next section. 

\section{A Plethora of Huge Operators and the Hoppe Model}\label{generalOperatorSection}

Now that we have seen hints of universality and commutativity in the context of a particular heavy operator insertion, let us try to generalize this. In the above analysis, a key simplification that made the problem tractable was that the heavy operator insertion depends only on one of the matrices, $X$. This allowed us to integrate out $Y$ analytically. 

In this section, we will pursue two lines of attack -- the first is analytic where we will consider more general heavy operators which are still functions of $X$ only. In particular, we will look at the following classes
\begin{itemize}
    \item Potentials: The simplest deformations of the Hoppe model is to add an interaction term of the form $\,\tr V(X)$ to the action or equivalently an insertion of $O_{\text{huge}}(X) = \exp( N\tr V(X))$. We will only consider single trace polynomial potentials $V(x) = \sum_k t_k x^k/k$.
    \item Classical Sources: We can couple the Hoppe model to a classical source by introducing a background hermitian matrix $J$ coupling to $X$ via the insertion $O_{\text{huge}}(X)=\exp( N\tr JX)$. We already saw an example of this in the previous section but now we will generalize to any source $J$.
    \item Characters or Schur Polynomials: These form a basis of $U(N)$ invariants of a single matrix. They are labeled by a Young tableau and when the tableau has $O(N^2)$ boxes, it is heavy enough to backreact.
\end{itemize}
Later, we will consider more general operators $O_{\text{huge}}(X,Y)$ depending on both matrices and we will be forced to resort to the second line of attack, Monte Carlo numerics. Let us now go through the above classes of huge operators one by one and convince ourselves that we have a universality, i.e. the expectation values of all probes in these backgrounds are fully determined by specifying a few numbers. 

\subsection{Potentials}

\begin{figure}
    \centering
    \includegraphics[width=0.8\linewidth,trim={0cm 5cm 0cm 0cm},clip]{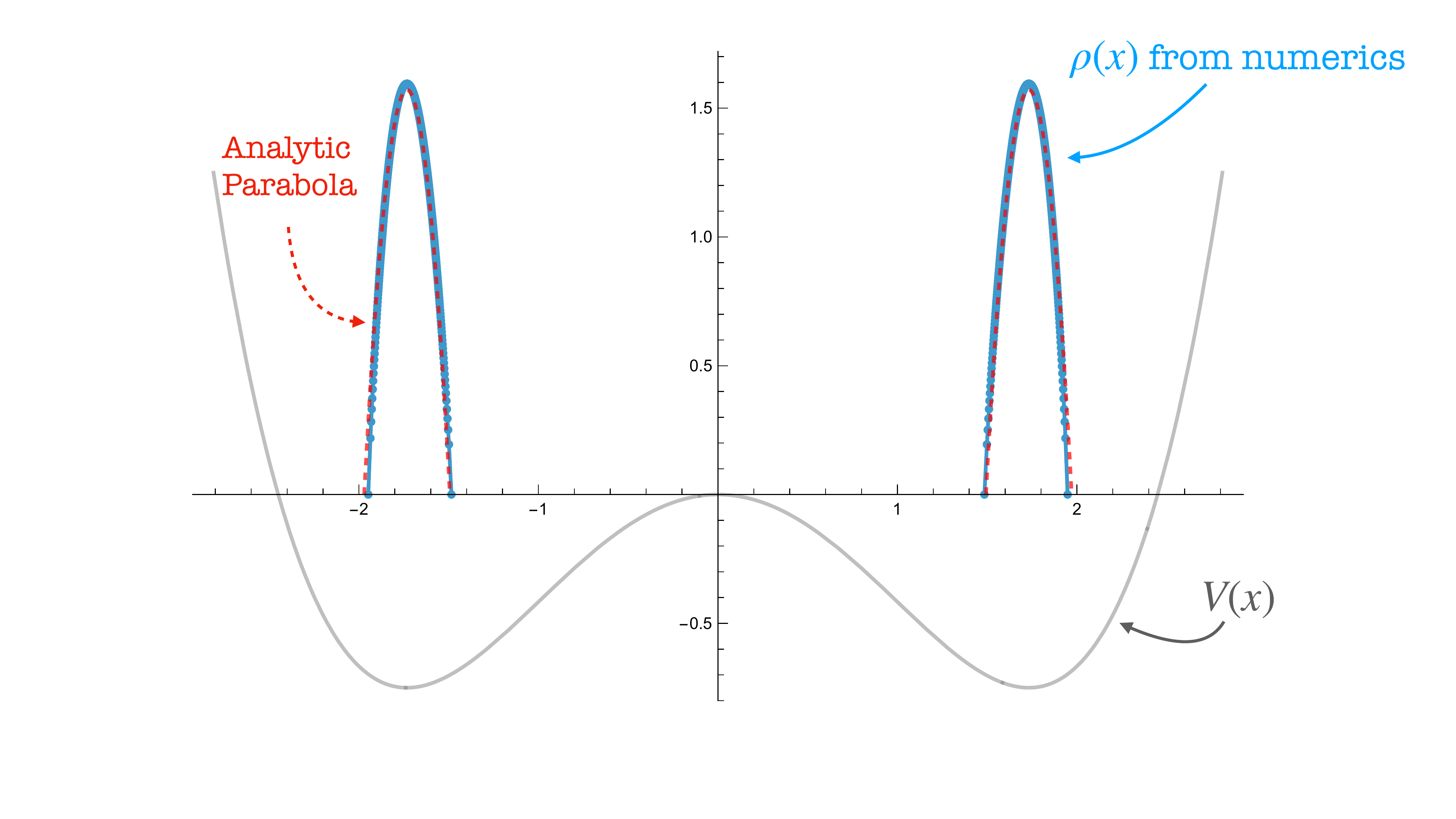}
    \caption{Eigenvalue density $\rho(x)$ for the quartic potential $V(x)=-\frac{x^2}2+g\frac{x^4}{4}$ at $g=\frac13$ and $\lambda=15000$. We show here the leading saddle with has filling fractions $\mathcal S_1=\mathcal S_2=\frac12$. The blue dots are from solving the discrete SPEs that follow from \eqref{spePotentialNoApprox} with $N=1000$. They agree well with the dashed red lines which are the analytic parabolas \eqref{densityMultiCut}.}
    \label{doubleWellFig}
\end{figure}

For the exponential potential operators after integrating out $Y$, we have
\beq
    \int \prod_i dx_i\, \Delta_\lambda(x)^2 \exp\left[-N\sum_{i=1}^NV(x_i)\right] \nn
\eeq
where we absorbed the Gaussian term into $V(x) = \sum_{k=0}^\infty\frac{t_k}{k} x^k$. In \cite{Kazakov:1998ji}, it was shown that the free energy for this integral at finite $N$ is given by a tau-function of the KP hierarchy. Here, we are only interested in the large $N$, large $\lambda$ limit. So let us proceed as before, and look at the saddle point equations
\beq
    2\fint dy\, \frac{\rho(y)}{(x-y)(1+2\lambda(x-y)^2)} - V'(x)=0\label{spePotentialNoApprox}
\eeq
At strong coupling \eqref{strongCouplingDensity},
\beq
    \rho'(x) = -\frac{\sqrt{2\lambda}}{2\pi}V'(x) + \ldots \label{speExponentialOps}
\eeq
Notice that for smooth potentials $V(x)$ at generic $x$, the RHS is very large. But the density must obey the normalization condition $\int dx\,\rho(x)=1$. Therefore, for the above equation to make sense, the support of $\rho(x)$ must be localized near the zero(s) of $V'(x)$. In general we can have multi-cut solutions with different filling fractions. Let's say there's only a single cut near $x=x_0$ which is a minimum of $V(x)$. Expanding around this point we have
\beq
    \rho'(x) = -\frac{\sqrt{2\lambda} V''(x_0)}{2\pi} (x-x_0) + \ldots \label{speExponentialOpsStrong}
\eeq
which is the same equation we encountered in solving the vacuum distribution of the Hoppe model but with a renormalized coupling. Therefore, $\rho(x)$ is again a parabola, centered at $x=x_0$ and whose size is given by 
\beq
    L_{\text{eff}} = \frac{(3\pi)^{\frac13}}{(2\lambda_{\text{eff}})^{\frac16}}\ , \qquad \lambda_{\text{eff}}=\lambda V''(x_0)^2 \nn
\eeq
We can easily extend this to the multi-cut case by noting that the SPEs reduce to \eqref{speExponentialOpsStrong} for each cut separately -- the ``off-diagonal" terms in \eqref{speExponentialOps} where $x$ and $y$ belong to different cuts are sub-leading. Therefore, as long as the potential is smooth, the distribution of eigenvalues is a union of parabolas, each with a center and size determined by the location of the extrema of the potential and the local curvature $V''(x_0)$. In this case, we need to further specify the filling fractions $\mathcal S_i$ (the fraction of eigenvalues that sit inside the $i^{\text{th}}$-cut). The density is given by,
\beq
    \rho(x) = \begin{cases}\displaystyle
        \frac{3\mathcal{S}_i}{4L_i^3} \left(L_i^2-(x-x^*_i)^2\right), \qquad\qquad & x\in [x^*_i-L_i, x^*_i+L_i]\\
        0 &\text{otherwise}
    \end{cases}\label{densityMultiCut}
\eeq
where $x=x^*_i$ are the locations of the critical points and
\beq
    L_i = \frac{(3\pi\mathcal{S}_i)^{\frac13}}{\left(2\lambda V''(x^*_{i})^2\right)^{\frac16}} \nn
\eeq
For non-generic potentials, for instance when the second derivative also vanishes, we get non-universal distributions. In figure \ref{doubleWellFig}, we plot the density for the quartic double well potential, $V(x) = -\frac{x^2}2+g\frac{x^4}{4}$ whose critical points are $x=0$ (a local maximum) and $x=\pm\frac1{\sqrt{g}}$ (the minima). The leading saddle turns out to be two cut i.e. two parabolas centered at the two minima as seen in the figure.\footnote{For an asymmetric double well with $V'(x)=g (x-a)x(x-b)$ with $a<0$ and $b>0$ we can evaluate the action for two parabolas: one centered at $a$ with filling fraction $\mathcal{S}$ and another centered at $b$ with filling fraction $1-\mathcal{S}$. We obtain $\log Z \simeq -N^2 S$ with $-S\simeq -\tfrac{1}{2} \log(\lambda)+f_1+f_2 \,\mathcal{S}$ with $f_1=f_1(a,b,g)$ and most importantly $f_2=f_2(a,b,g)=(\texttt{positive quantity}) \times  (a+b)$. The filling fraction dependence is a trivial ramp whose sign only depends on the sign of $a+b$. If $b+a$ is positive/negative the preferred configuration is to have all eigenvalues in the left/right cut  with $\mathcal{S}=1/0$ respectively. This makes intuitive sense since $V(a)$ is smaller/bigger that $V(b)$ when $b+a>0$/$b+a<0$ respectively. When the well is symmetric with $b=-a$ the ramp becomes horizontal and all saddles are degenerate to leading order. An analysis of the next order would show that $\mathcal{S}=1/2$ is the dominant saddle as described in the text. Of course, in any case, we are often interested in leading as well as sub-leading saddles as well. After all, at large $N$, tunneling between these saddles is suppressed so we could very well ``live" in one of those metastable vacua.}

Another simple, albeit non-polynomial potential we could consider is 
\beq
    \tr V(X) = K \Tr \log X \label{detPot} \,.
\eeq
This potential is equivalent to inserting a stack of $K$ determinants, $\det X^K$ which is a huge operator when $K\sim N$. In this case, the above computation goes through and we have two minima at $x^*=\pm\sqrt{K/N}$ and the density is given by \eqref{densityMultiCut}. The leading saddle has two-cuts with $\mathcal S_1=\mathcal S_2=\frac12$ and the numerics match these parabolic densities as seen in figure \ref{YTexamples}.
\subsection{Sources} \label{sourcesSec}

Let us turn to the next class of huge operators -- classical sources $e^{N\tr J X}$. At first glance, this seems similar to the above case of single trace polynomial potentials however, the source $J$ is a free matrix parameter which makes these operators much more general. For instance, as we saw for the homogeneous line source, the effective potentials generated by this source is non-polynomial and includes all multi traces \eqref{homoLineVandermondeRatio}. Even at zero coupling, the analytic solution is only known for some special cases like the cubic potential (Kontsevich model) solved by Gross-Newman \cite{gross:1991,Gross:1991ji}. For general source $J$ at non-zero coupling, recall from section \ref{introSection} that the integral we are interested in is
\beq
    \int \prod_i dx_i\, \Delta_\lambda(x)^2 \exp\left[-\frac N2\sum_i x_i^2\right] I(x, j) \label{Iintegral}
\eeq
where $\{j_i\}$ are the eigenvalues of the source and $I(x,j)$ is the HCIZ integral,
\beq
    I(x,j) = \int dU\ e^{N\tr XJ} = \frac{\det\limits_{1\leq a,b\leq N} e^{Nj_ax_b}}{\Delta(j)\Delta(x)}\times \mathfrak{N} \nn
\eeq
where $\mathfrak{N}$ is a trivial normalization constant defined in \eqref{hcizFiniteN}. Unlike the case of a homogeneous source, the saddle point equations for the eigenvalues involve complicated derivatives of the $N\times N$ determinant above. It turns out that HCIZ integral has a very nice large $N$ limit in terms of a one-dimensional fluid flow \cite{Matytsin_1994}, with rich connections to integrable models that we can leverage. We review some aspects of this connection in appendix \ref{fluidAppendix}. The key facts that we need here are that the HCIZ integral is computed by a one-dimensional integrable fluid flow described by the density and velocity functions $\rho(x,t)$ and $v(x,t)$ which obey 1d Navier-Stokes equations \eqref{HCIZfluidFlowEquations}. The fluids obeys the following boundary conditions
\beq
    \rho(x,t=0) = \texttt{density of }\{x_i\}\qquad \text{and} \qquad \rho(x,t=1) = \texttt{density of }\{j_i\} \label{bc1}
\eeq
Furthermore, the saddle point equations for the $x$ eigenvalues in (\ref{Iintegral}) yield the following gluing condition at $t=0$,
\beq
    \fint dy\,\rho(y,t=0)\left(\frac1{x-y} - \frac{4\lambda (x-y)}{1+2\lambda(x-y)^2}\right) + v(x,t=0) = 0\label{bc2}
\eeq
Now, what we are after is the density of $x$'s, $\rho(x,0)$ given the density of $j$'s, $\rho(x,1)$ at strong coupling $\lambda\rightarrow\infty$. In principle we need to solve the fluid flow equations \eqref{HCIZfluidFlowEquations} together with the boundary conditions \eqref{bc1} and \eqref{bc2}. However, as alluded to above, the fluid flow is integrable and there exist conserved charges which relate densities and velocities at $t=0$ with those at $t=1$. In order to write these down, it is convenient to define the functions
\beq
    G_+(x) = x+v(x,0)+i\pi \rho(x,0) \ , \qquad G_-(x) = x-v(x,1)-i\pi \rho(x,1) \nn
\eeq
which are defined for $x$ above the cuts and then one can analytically continue to the whole complex plane. In terms of these functions, there exist the following conserved charges
\beq
    \frac1{m+1}\oint \frac{dx}{2\pi i}\, x^n G_+(x)^{m+1} = \frac{-1}{n+1} \oint \frac{dx}{2\pi i}\, x^m G_-(x)^{n+1} \nn
\eeq
Inspired by the above observations of universality, we make a parabolic ansatz for the density centered at some $x_0$ with a size $L$. Plugging this ansatz into \eqref{bc2}, we can solve for the initial velocity in terms of $x_0$ and $L$. We can then use the first two conserved charges to fix the parameters to be (see appendix \ref{fluidAppendix} for details)
\beq
    x_0 = \Tr J\ , \qquad L=L_{\text{vacuum}} \times \frac1{\big(1-\Tr J^2 + (\!\Tr J)^2\big)^{\frac13}}\label{parabolaForJX}
\eeq
This result reduces to \eqref{LengthAlpha} for the homogeneous source where $\Tr J=0$ and $\Tr J^2={\alpha^2}/{12}$. In particular when $J=0$, we go back to the vacuum parabola. Just like in the homogeneous case, we have a divergence in $L$ when the variance of the eigenvalues of $J$ becomes equal to one. Beyond this point, we have a different, non-universal phase.

An interesting perspective on the classical source operators described in this section is that they can be thought of as ``Coherent State operators", $\int dU e^{N\tr JUXU^\dagger}$ introduced in \cite{Berenstein:2022srd} where they study generating functions for BPS operators in $\mathcal{N}=4$ Super Yang-Mills. When we insert a coherent state operator into the Hoppe model, the unitary integral can be absorbed into $X$, so it is equivalent to having a classical source. This is of course not true if we insert multiple coherent state operators. It would be interesting to explore what happens in that case.

\subsection{Characters}

\begin{figure}[t]
    \centering
    \includegraphics[width=\linewidth,trim={0cm 15cm 0cm 0cm},clip]{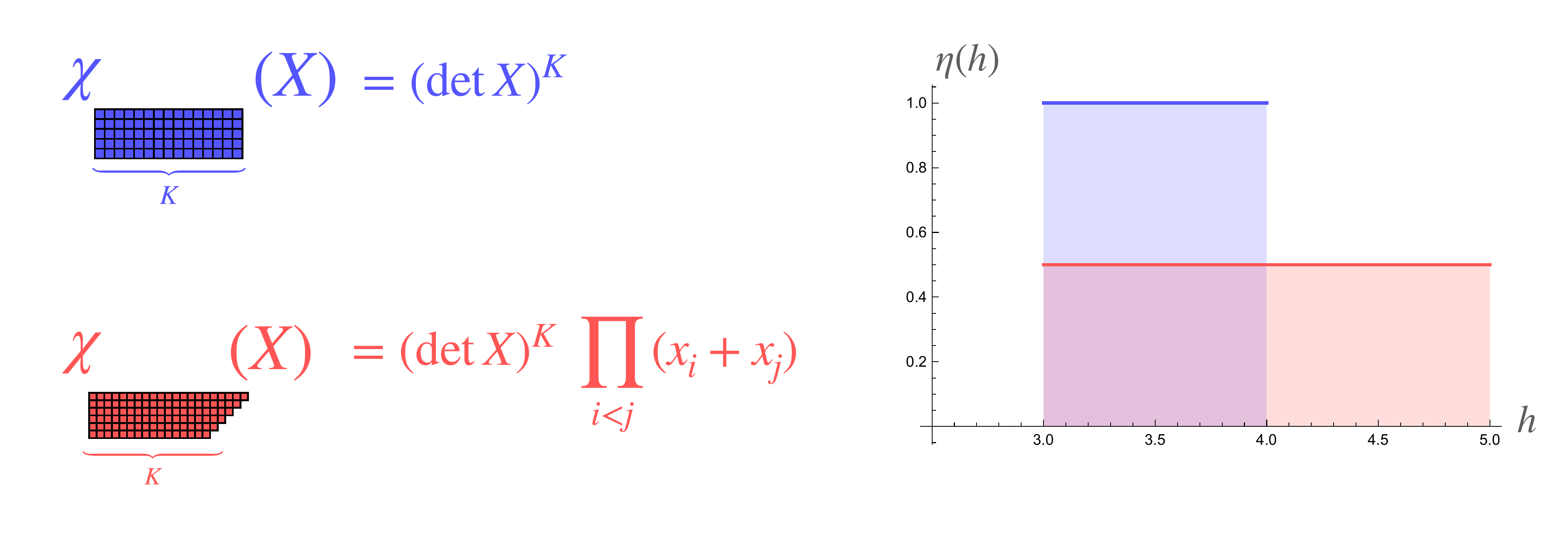}
        \includegraphics[width=0.9\linewidth,trim={0cm 15cm 0cm 0cm},clip]{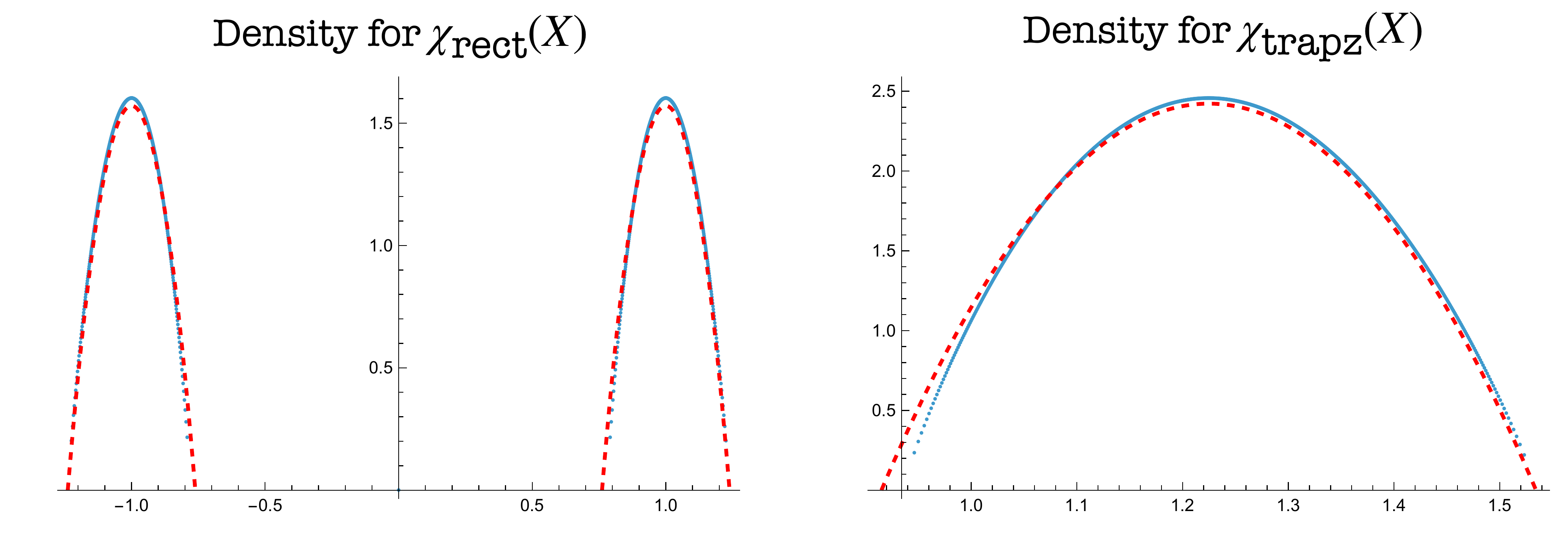}
    \caption{{\bf Top:} Two simple examples of huge Character operators corresponding to a rectangular YT (in blue) and trapezium YT (in red). A rectangle with each row having $K$ boxes corresponds to a stack of $K$ determinants and a trapezium with $i^{\text{th}}$ row having $K+N-i$ boxes corresponds to a stack of determinants times a vandermonde-like factor (with sums instead of differences). At large $N$, we can define a density of the shifted highest weight $h_i$ and for the rectangle, it is a uniform distribution on the interval $\left[\tfrac KN,\tfrac KN+1\right]$. For the trapezium, it is also a uniform distribution but on the interval $\left[\tfrac KN,\tfrac KN+2\right]$. {\bf Bottom:} Again, we find universal parabolas for these huge operators as described in the main text.}
    \label{YTexamples}
\end{figure}

Let us now consider the last class of huge operators -- Characters $\chi_R(X)$. These are labeled by a Young tableau $R$ with $N$ rows and are defined as follows
\beq
    \chi_R(X) = \frac{\det\limits_{1\le i,j\le N} x_i^{h_j}}{\Delta(x)} \nn
\eeq
where we introduced the shifted highest weight $h_i = R_i +N-i$ where $R_i$ is the number of boxes in the $i^{\text{th}}$ row of the Young tableau. These are manifestly $U(N)$ invariants since they are symmetric polynomials of the eigenvalues. A nice feature of these operators is that they form a basis of $U(N)$ invariants of a single matrix $X$. Two simple examples of these characters for the rectangular Young Tableau and trapezium Young tableau are shown in figure \ref{YTexamples}. We also show the density of the highest weights at large $N$ in the figure.

We want to now consider the Hoppe model with a~$\chi_R(X)$ insertion where the Young tableau~$R$ has~$O(N^2)$ boxes. When the coupling $\lambda$ is turned off, the integral can be done in closed form at finite $N$ \cite{Kazakov:1995ae} and there is no universality -- everything depends on the Young tableau $R$. At strong coupling, there exists a universal, black-hole regime. In order to see this, we can once again use the HCIZ fluid flows due to the following observation
\beq
    \chi_R(X) = I(\log x, h) \frac{\Delta(\log x)\Delta(h)}{\Delta(x)\mathfrak{N}}\,\label{characterAsHCIZ}
\eeq
where $\mathfrak{N}$ is a normalization constant defined in \eqref{hcizFiniteN} that does not affect the saddle point equations. Following the same steps as above -- write down boundary conditions for the fluids, make a parabolic ansatz and solve for the parameters using conserved charges -- we obtain
\beq
    x_0 = \sqrt{\langle h\rangle -\frac12}\ ,\qquad L=L_{\text{vacuum}} \left(\frac{6(2\langle h\rangle-1)}{1+12(\langle h\rangle^2+2\langle h\rangle-\langle h^2\rangle-1)}\right)^{\frac13}\label{parabolaParamsCharacter}
\eeq
where $\langle h^n\rangle$ is the $n^{\text{th}}$ moment of the distribution of highest weights $\{h_i\}$. See appendix \ref{fluidAppendix} for details of the derivation. Note that when $h_i=K+N-i$, i.e. for the rectangular Young tableau, the character is a stack of determinant $\chi_R(X) = \det X^K$. Plugging in this distribution we get $x_0 = \sqrt{K/N}$, in agreement with the discussion below \eqref{detPot}. In figure \ref{YTexamples}, we plot the densities for the leading saddles of the rectangle (two cut) and the trapezium (one cut) and compare them with the analytic parabolas finding again a perfect match.


\subsection{Huge Operators $O(X,Y)$ -- Hybrid Monte Carlo}

\begin{figure}[t]
    \centering
    \includegraphics[width=\linewidth,trim={0cm 8cm 0cm 0cm},clip]{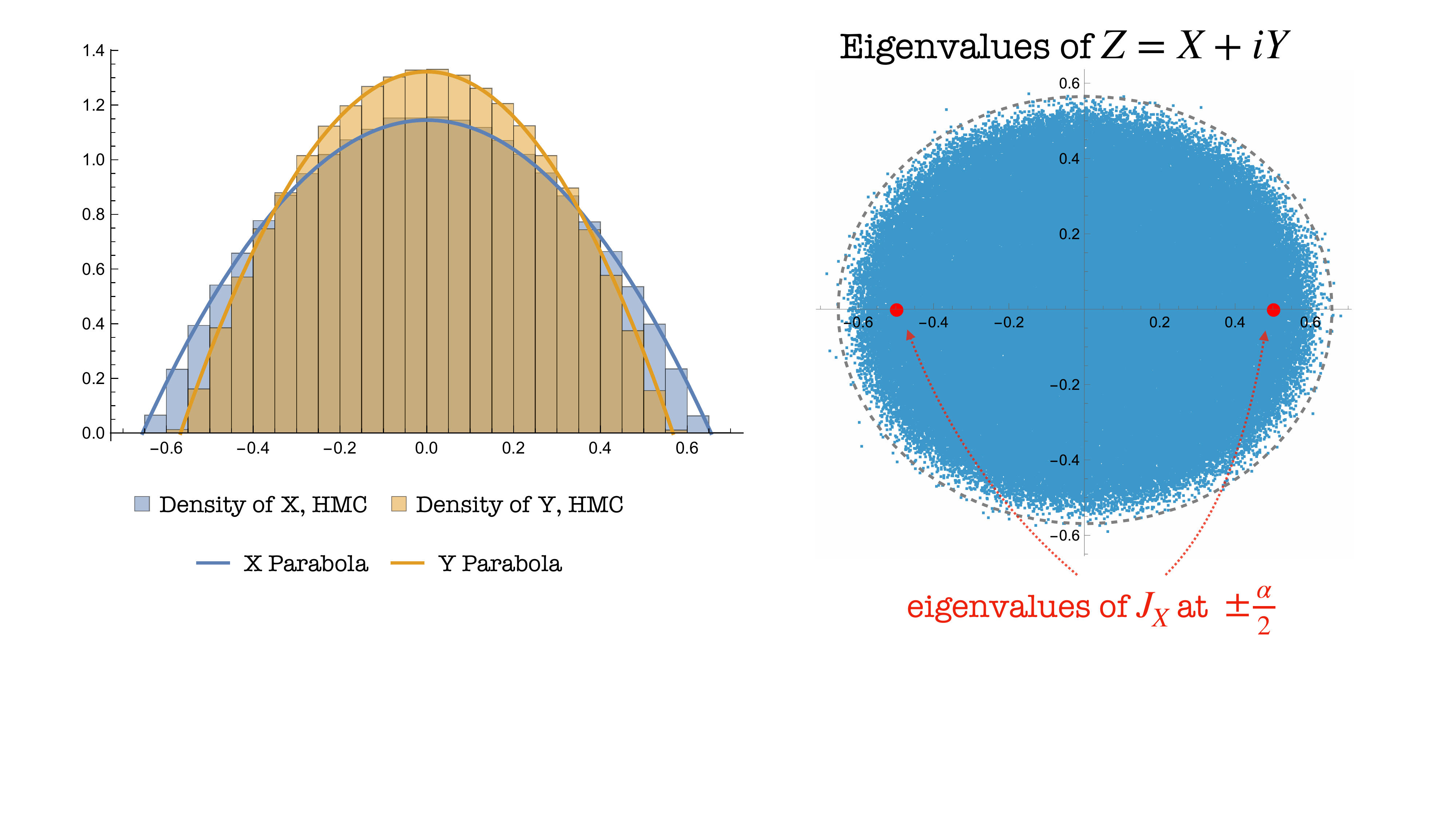}
    \caption{HMC results for the source \eqref{twoCenterSourceEq} at $\alpha=1<\alpha_c$. These results are at $N=200$ and $\lambda=1000$. On the left, we plot histograms of eigenvalues of $X$ and $Y$ which agree with the analytic parabolas where $L_X$ is obtained from \eqref{parabolaForJX} and $L_Y$ from \eqref{LyLx}. On the right, we show a point cloud of the eigenvalues of the matrix $Z=X+i Y$ (for $1000$ steps of the HMC Markov chain) and also the location of the source's eigenvalues in red. Note that at strong coupling the matrices commute and the real and imaginary parts of eigenvalues of $Z$ are precisely the eigenvalues of $X$ and $Y$ and their joint density is given by \eqref{XYJointEllipse} whose support is shown here as the the dashed gray ellipse.}
    \label{twoCenterFig}
\end{figure}

The skeptical reader might complain that the restriction to huge operators that are independent of $Y$ is too strong and that the universality would break down once we relax this requirement. Indeed, we do not know how to make analytic progress when we cannot integrate out $Y$. So we resort to numerical Monte Carlo simulations for these cases. We use the Hybrid Monte Carlo (HMC) method which is very friendly to large $N$ matrix integrals that have no sign problem (like the Hoppe model). A useful reference for us was the very nice review \cite{Jha:2021exo} by Raghav Jha with ready to use Python code which we took and adapted; one other review we found very instructive is \cite{Hanada:2018fnp}.
We discuss the details of HMC in appendix \ref{app:HMC}. An important step in the HMC is the computation of the ``force" i.e. ${\partial S}/{\partial X_i}$ which is particularly simple when we have classical sources for $X$ and $Y$,
\beq
    \int dX\,dY\ \exp\left[N\tr\left(-\frac12 X^2 - \frac12 Y^2 + \lambda [X,Y]^2 + J_X X + J_Y Y\right)\right] \nn
\eeq
where $J_X$ and $J_Y$ are some fixed external matrices. We tried various choices for them and once again find that there exists a universal black-hole regime where the eigenvalue distributions are parabolic. Let us look at some examples.

\begin{figure}[t]
    \centering
    \includegraphics[width=0.95\linewidth]{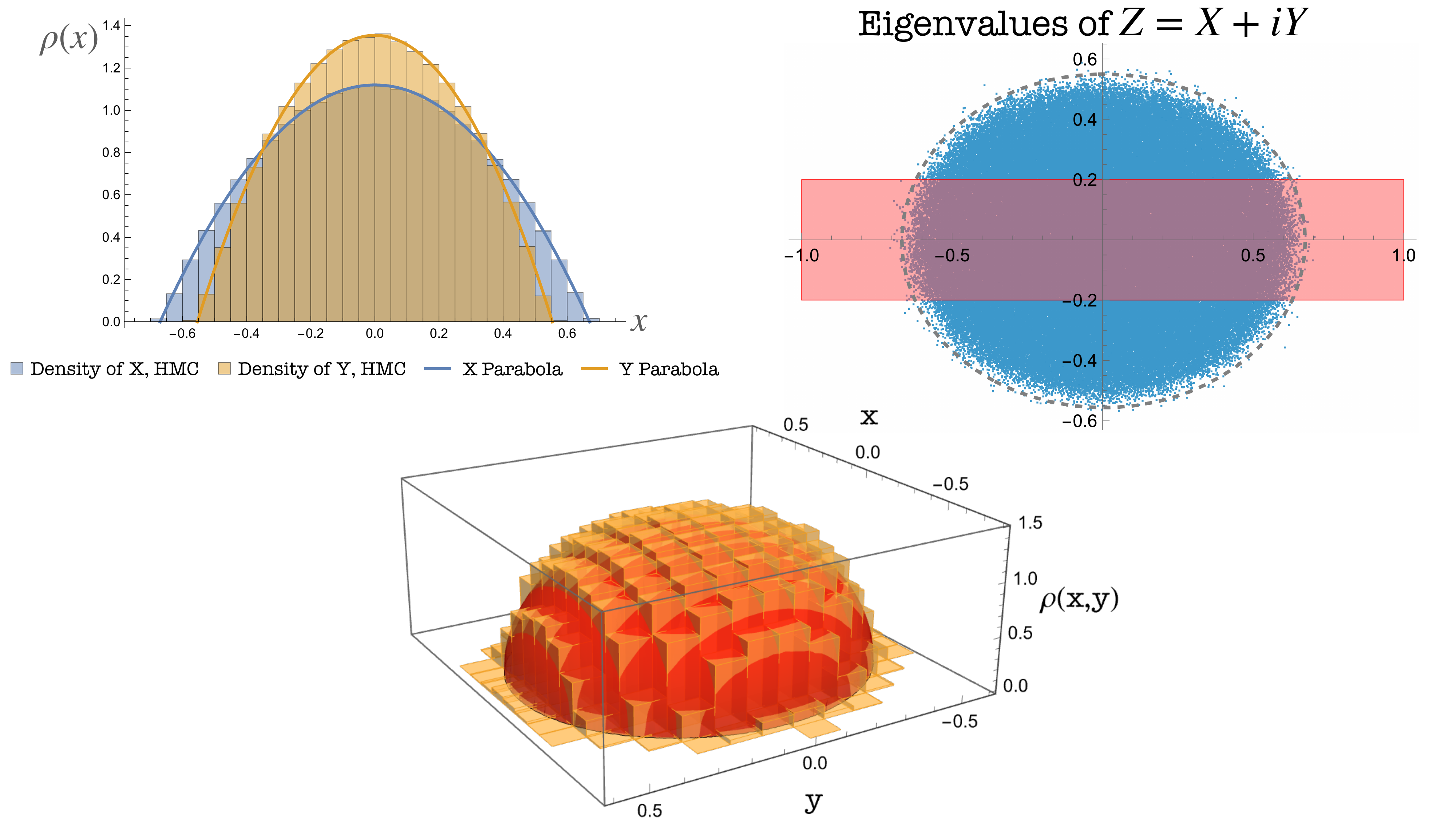}
    \caption{Results from HMC for the case of a rectangular source \eqref{rectangleSource} with $\alpha=2$ and $\beta=\frac25$ shown as the red rectangle in the top-right panel. For these parameters, we are in the universal regime. The $1$d and $2$d eigenvalue distributions are shown here and they match perfectly with the universality predictions (the solid lines for $1$d and the red ellipsoid for $2$d).}
    \label{rectanglePreCrit}
\end{figure}

As a warm-up, we can try to confirm the analytic results we found above when $J_Y=0$. Consider for example
\begin{equation}
    J_X = \text{diag}\Bigl(\underbrace{-\frac\alpha2,-\frac{\alpha}2\ldots }_{N/2 \text{ times}},\underbrace{+\frac\alpha2,\ldots +\frac\alpha2}_{N/2 \text{ times}}\Bigr)\ \qquad \text{and}\qquad J_Y=0\label{twoCenterSourceEq}
\end{equation}
where $\alpha>0$. Note that we cannot solve this with eigenvalue numerics because we have no simple identity like \eqref{homoLineVandermondeRatio} for this source. From the analytic result \eqref{parabolaForJX}, we know that there should be a phase transition when $\Tr J_X^2 - \left(\Tr J_X\right)^2=1$ which corresponds to a critical $\alpha_c = 2$. We expect a parabolic density for $\alpha<\alpha_c$, which is confirmed by figure \ref{twoCenterFig}~\footnote{Interestingly, in this case even after the phase transition $\alpha>\alpha_c$, the distribution is given by two parabolas, centered at $\pm \frac\a2$}. 


Now, let us consider more general sources with both $J_X$ and $J_Y$ non-zero. For instance, let's take the eigenvalues of $J_X$ and $J_Y$ to be uniformly distributed inside a rectangle,
\begin{equation}
    \begin{aligned}
        J_X = \diag{j_X}\ \ &, \qquad J_Y=U^\dagger \diag{j_Y}U\\
        (j_X, j_Y)&\in \left[-\frac\alpha2,\frac\alpha2\right]\times \left[-\frac\beta2,\frac\beta2\right]
    \end{aligned}\label{rectangleSource}
\end{equation}
where $U$ is a random unitary and $\alpha$ and $\beta$ are the side lengths of the rectangle. The 1d eigenvalue distribution $\rho(x)$ is again a parabola with $L=O(\lambda^{-\frac16})$ when $(\alpha, \beta)$ are small enough. We don't have an analytic prediction for the size of the parabola, but we can infer this by measuring the moments with HMC
\begin{equation*}
    \langle\Tr X^2\rangle=\frac{L_X^2}5\ \, \qquad\langle\Tr Y^2\rangle=\frac{L_Y^2}5 \nn
\end{equation*}
In figure \ref{rectanglePreCrit}, we plot the eigenvalue densities from HMC for $(\alpha,\beta)=(2,\frac25)$ and compare with the parabolas and ellipsoids obtained this way -- they are in agreement.

\begin{figure}[t]
    \centering
    \includegraphics[width=0.98\linewidth]{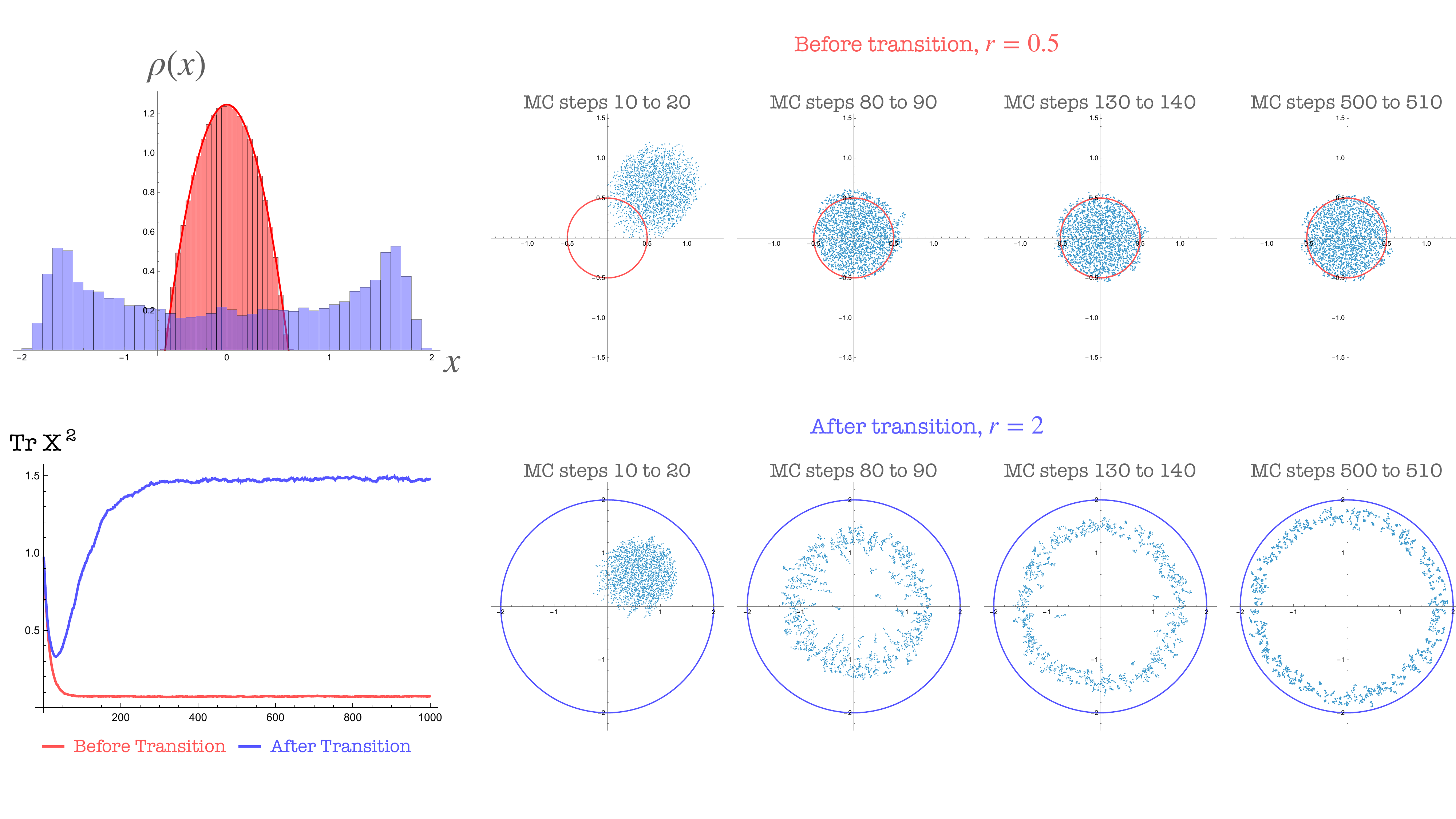}
    \caption{HMC results for ring sources where $J_X$ and $J_Y$ are diagonal and $J_X^2 + J_Y^2=r^2$. Here we consider two cases, $r=0.5$ (shown in red) and $r=2$ (shown in blue) which are before and after the phase transition as indicated by the densities shown on the left -- the solid red curve is a parabola fit. We initialize both HMC runs at the same point with the same hyper-parameters. As seen from the plot of $\langle\Tr X^2\rangle$, the universal case thermalizes faster than the non-universal one. The MC acceptance rate is also much higher for the universal case ($97\%$ vs $73\%$).}
    \label{fig:ringSource}
\end{figure}

We expect there to be a critical line $\alpha_c(\beta)$ in the $\alpha$-$\beta$ plane below which we have the universal phase. When $\beta=0$, we know from \eqref{LengthAlpha} that $\alpha_c(0)=\sqrt{12}$. One could try to numerically find the shape of this curve -- we did not do this here. When the source is above the critical line, the distribution is something non-universal with $L=O(1)$. 

We could keep going and pick eigenvalues of $J_X$ and $J_Y$ to be distributed in various ways. We tried a uniform disc, uniform triangle, uniform ring and so on. In all these cases, there exists a regime of universality. For general sources, we don't know how to predict the distributions a priori. The equation of an ellipse is a quadratic form with five moduli (we can think of them as the position of the center, major and minor axis lengths and orientation). So, once we measure say $\langle\!\Tr X\rangle,\langle\!\Tr Y\rangle,\langle\!\Tr X^2\rangle,\langle\!\Tr Y^2\rangle$ and $\langle\!\Tr XY\rangle$, we know everything there is to know about probes in these backgrounds at leading order. In general if we have $n$ ellipses, we would need $5n$ measurements to determine everything. 

\subsection{Phase Transitions and Exotic Non-Universal Objects}
If the potential is not too shallow; if the classical source is not too large; if the character is not too degenerate (e.g. too thin, see next section) we are in a universal phase. The one dimensional distribution of eigenvalues are universal parabolas and the two dimensional distributions are universal hemispheres. 

\begin{figure}[t]
    \centering
    \includegraphics[width=\linewidth,trim={0cm 5cm 0cm 0cm},clip]{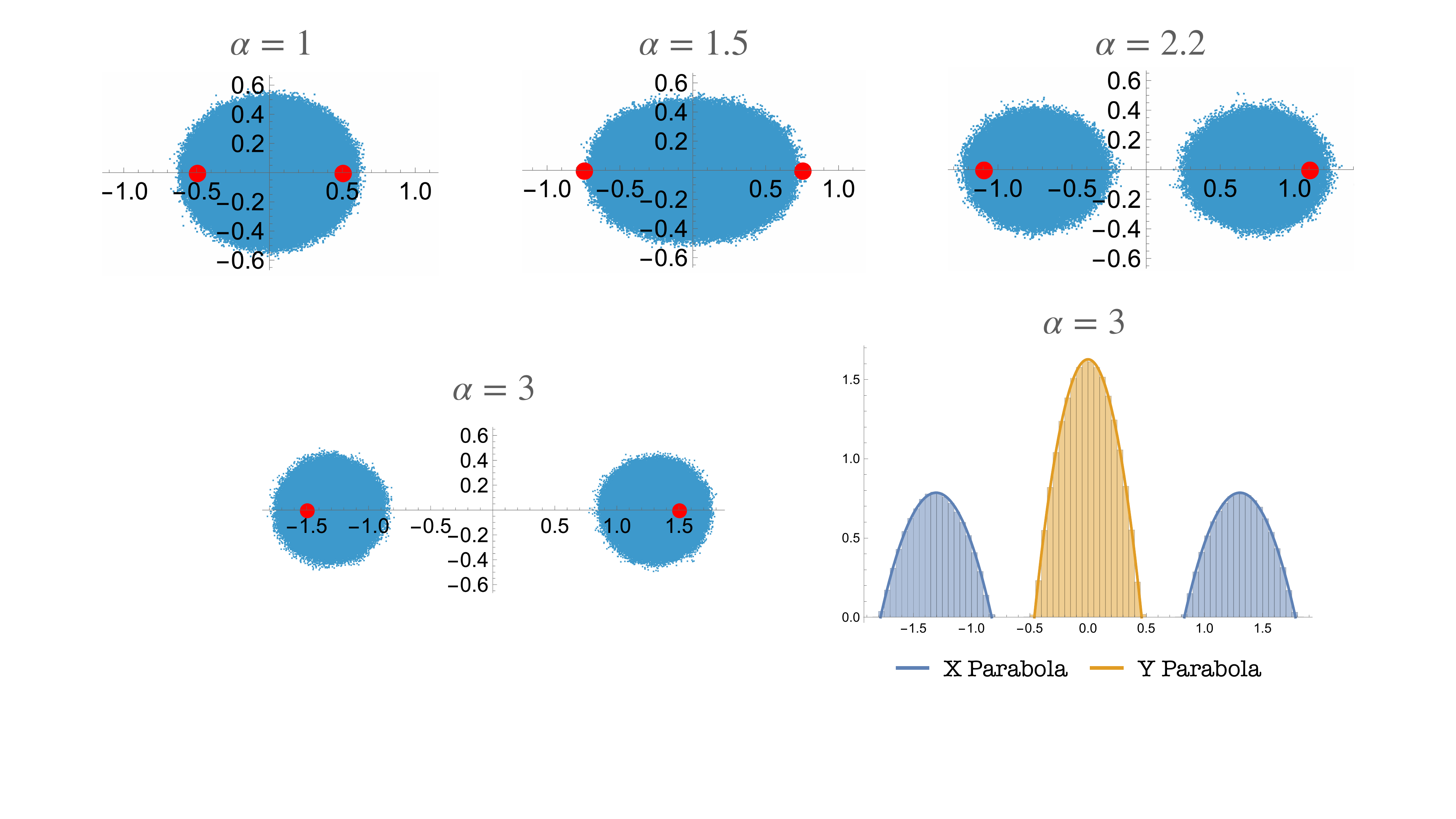}
    \caption{Phase transition for the source with two centers \eqref{twoCenterSourceEq}. The position of the source eigenvalues is shown as red dots and the blue points are eigenvalues of $Z=X+i Y$ as usual. As we increase $\alpha$, the single parabola phase ceases to exist at a critical $\alpha_c=2$. However, as seen in the bottom plot, the distribution after the phase transition is still a pair of parabolas for $X$ and single parabola for $Y$! (Note that a single center where $J= \alpha\,\texttt{Identity}$ is still the vacuum since we can simply absorb it by shifting $X$; It is thus not surprising if two centers become too separated we end up with half of the eigenvalues in each of the vacua.) }
    \label{fig:two-centers}
\end{figure}

\begin{figure}
    \centering
    \includegraphics[width=\linewidth,trim={0cm 12cm 0cm 0cm},clip]{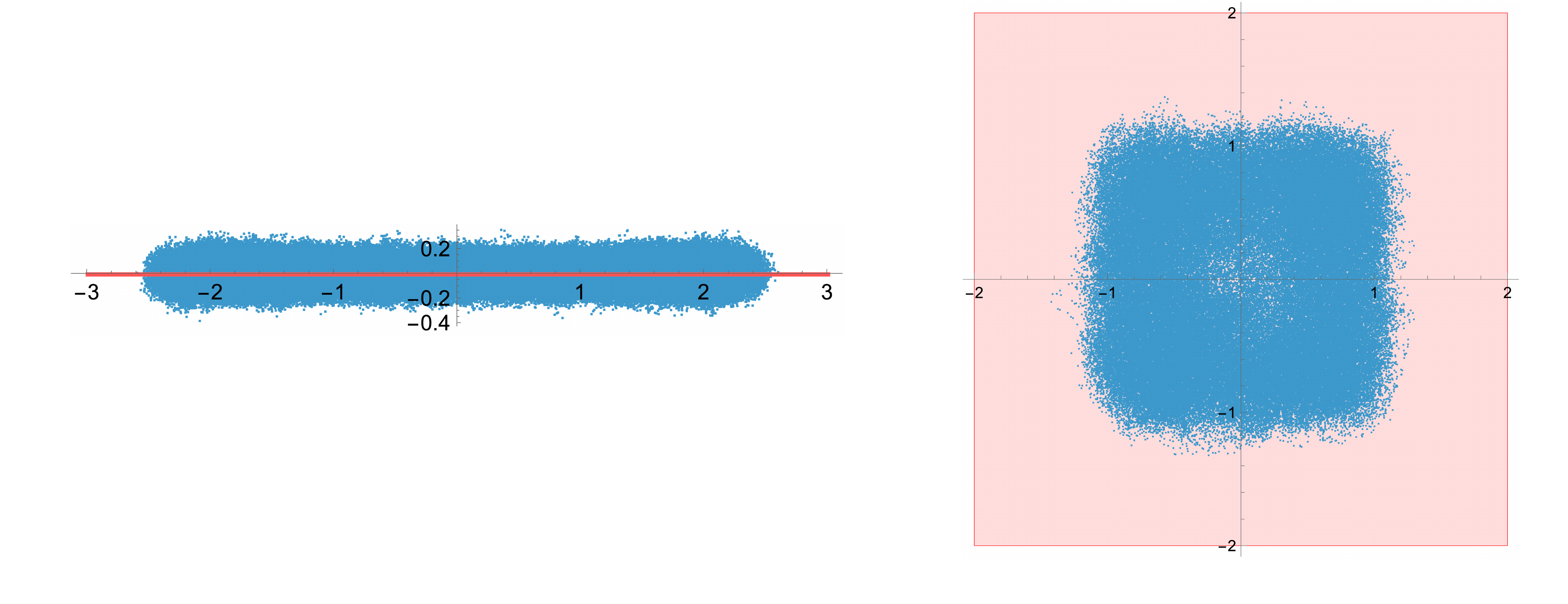}
    \caption{Here we show two more examples of non-universal distributions after the phase transition. On the left, we have a homogeonous source at $\alpha=6>\alpha_c$ and on the right, we have a uniform square source~\eqref{rectangleSource} with $\alpha=\beta=4$. The eigenvalues of $X$ and $Y$ are attracted by the eigenvalues of $J_X$ and $J_Y$ as seen in these plots. Different $J$'s give different densities.}
    \label{nonUnivFig3}
\end{figure}

Then we have phase transitions. When sources are bigger than some critical value we transition to a non-universal result. (That some transition ought to be there is clear, at least for commuting sources; After all, for huge sources we know the distribution of eigenvalues approaches that of the sources and are thus clearly no longer sharply parabolas, see section \ref{largeSourcesSec}.)  If the source is unidimensional we predicted analytically (in section \ref{sourcesSec}) that the phase transition occurs when $\Tr(J^2)-\Tr(J)^2$ reaches $1$.
For two dimensional sources we no longer have an analytic prediction but we can neatly observe this transition. We observed it in several examples and picked a few cases to illustrate here, see figures \ref{fig:ringSource}, \ref{fig:two-centers} and \ref{nonUnivFig3}. An interesting numerical observation we made is that HMC numerics are much harder in the non-universal regime than in the universal regime. We need to pick smaller time-steps in the MC jumps in order to have a reasonable acceptance rate. This suggests that the universal saddles are smoother and the non-universal ones are very sharply peaked in phase space. $\diamond$ It would be nice to have a notion of entropy for these saddles such that the universal ones -- like black holes -- have much more entropy than the non-universal ones $\diamond$.


\section{Abelianization}
This section is outside the main scope of this paper. Here we describe a very different class of huge operators. These huge operators are very different because despite being huge -- that is made out of a huge number of fields -- they do not change the  density $\rho(x)$ of eigenvalues $(x_1,\dots,x_N)$ in a smooth way. The simplest class of such operators, for example, leaves all eigenvalues untouched except for a single one, say $x_1$ which becomes parametrically large and decouples from the other ones. This decoupling of one of the eigenvalues such that $U(N)\to U(1)\times U(N-1)$ is what we call \textit{Abelianization}. We can also encounter richer symmetry breaking patterns where $U(N)\to U(2)\times U(N-2)$ and so on. Since the distribution of most eigenvalues is untouched in these examples, we would \textit{not} say that the background is deformed but instead that a very heavy object -- parametrized by the location of the eigenvalues which decoupled -- was added in our original background. In holographic examples, this could be heavy very large D-branes added on top of some geometry as in \cite{Yamaguchi:2006te,Fraser:2011qa}. What we point out in this section is that this phenomena is very general and can be observed already in the simplest possible matrix models, including the one matrix model. To our knowledge, this is a new observation. We will see that these huge operators will exhibit their own very interesting type of universality.

The simplest possible example we want to consider is this single matrix integral
\beq
Z_{{\color{red}p},{\color{blue}q}}= \int dX\!\!\!\!\!\!\!\!\underbrace{(\Tr X^{\color{red}p})^{\alpha N^2/{\color{red}p}}}_\texttt{Huge operator with $\alpha N^2$ X's} \!\!\!\!\!\!\exp\Big[\underbrace{-N \tr \Big(\frac{t_2}{2}  X^2 + \frac{t_3}{3} X^3+ \dots  + \frac{t_{\color{blue}q}}{{\color{blue}q}}X^{\color{blue}q}\Big)}_\texttt{Polynomial action with maximum degree {\color{blue}q}}\Big] \label{Zabelian}
\eeq
The phenomena we will observe will take place whenever $p>q$; in particular, it is there already for the simplest $q=2$ Gaussian case. 

The eigenvalue saddle point equations read
\beq
-t_2 x_i-\dots - t_{\color{blue}q} x_i^{{\color{blue}q}-1} + \alpha N \frac{x_i^{{\color{red}p}-1}}{x_1^{{\color{red}p}}+\sum\limits_{j=2}^N x_j^{{\color{red}p}}}+\frac{1}{N} \sum_{j\neq i} \frac{1}{x_i-x_j}=0 \,, \qquad i =1,\dots,N\,. \label{saddlePoint}
\eeq
In the denominator we simply split $\tr(X^p)=\sum_{j=1}^N x_j^{{\color{red}p}}$ into two terms for convenience.
We will now discuss two saddle solutions to these equations. 
\subsection{The Abelian Saddle: Same Background with a Huge Object Therein}
These saddle point equations (\ref{saddlePoint}) admit a simple solution where $x_1$ is much larger than all other eigenvalues and their sums. When this is the case, the equation for $i=1$ simplifies to just the two middle terms, 
\beq
-t_{\color{blue}q} x_1^{{\color{blue}q}-1} + \frac{\alpha N}{x_1} \,\, \Rightarrow \,\, x_1 =  \left(\frac{\alpha N }{t_{\color{blue}q}}\right)^{1/{\color{blue}q}} \equiv x_* \,. \label{x1Eq}
\eeq
Moreover, for the $i>1$ equations we can drop the source term to get 
\beq
-t_2 x_i-\dots - t_{\color{blue}q} x_i^{{\color{blue}q}-1} +\frac{1}{N} \sum_{j\neq i,1} \frac{1}{x_i-x_j}=0 \,, \qquad i =2,\dots,N\,. \nn
\eeq
which are nothing but the vacuum equations without the eigenvalue $x_1$! We conclude that for this saddle, $(x_2,\dots,x_N)$ follow the very same distribution as the vacuum without the huge operator while a single eigenvalue $x_1$ decouples from the group and scales as $N^{1/{\color{blue}q}}$. We can say that the background (the $x_2,\dots,x_N$) is not deformed but there is an extra heavy object on top of that background (the $x_1$). Note that this is a consistent approximation if indeed we can drop~$\sum_{j=2}^N x_j^{{\color{red}p}} =O(N)$ compared to~$x_1^{{\color{red}p}}=O(N^{{\color{red}p}/{\color{blue}q}})$ which requires $$p>q$$ as anticipated; only then do we obtain (\ref{x1Eq}). Moments of light operators in the presence of this huge operator will be dominated by the huge eigenvalue so that 
\beq
\langle \tr X^n \rangle \simeq (x_*)^n\,. \label{xstarmoments}
\eeq
for any $n>q$. For $n=q$ the bulk eigenvalues also contribute and the result is the sum of that semi-circle contribution plus this. For $n<q$ only the bulk eigenvalues contribute. We find this amusing. Some probes for $n>p$ are universal and decouple from the background while others do not care about $x_*$ at all. $\diamond$ An analogy could be a space-time geometry like a black-hole with a simple asymptotic region like flat space. A huge object like a very heavy brane very far from the black hole would not feel the black hole at all while some observers near the black hole would not care about the far away heavy object $\diamond$. 

The result (\ref{xstarmoments}) is extremely simple with all moments governed by the single location~$x_*$. Moreover they do not depend on $p$ and in fact only depends on the highest $x^q$ power of the polynomial potential which enters in defining $x_*$ in (\ref{x1Eq}). In this sense it is yet another sort of universality. Note that in contradistinction with the usual large $N$ scaling in more conventional t'Hooft expansions, here different $n$'s in (\ref{xstarmoments}) have a different scaling with $N$.

As far as we know, the Abelianization mechanism we just described had not been highlighted before as a general phenomena in the context of these simplest one matrix models. But very similar instances of single eigenvalues decoupling from the bunch and with and interpretation in terms of D-brane locations and so on did appear before in richer holographic models, see for example the nice discussions around figure 2  in \cite{Fraser:2011qa}, figure 3 in \cite{Yamaguchi:2006te} or formula 6.41 in \cite{Choi:2024xnv}; see also \cite{Hartnoll:2006is}.

\subsection{A More Conventional Saddle: A New Background}
There is a more conventional solution of the saddle point equations. Denoting 
\beq
\mathbb{C} \equiv \Tr(X^p)=\frac{1}{N} \sum_{i=1}^{N} x_i^p \,, \nn
\eeq
the saddle equations (\ref{saddlePoint}) simply read 
\beq
-V_\text{eff}'(x_i) +\frac{1}{N} \sum_{j\neq i} \frac{1}{x_i-x_j}=0 \,, \qquad V_\text{eff}(x)=\sum_{k=1}^{{\color{blue} q}} \frac{t_k}{k} x^k + \frac{\alpha}{\mathbb{C}} \frac{x^{\color{red} p}}{{\color{red} p}} \,.\label{saddlePoint2}
\eeq
One possibility is to solve these equations as one would in a standard matrix model for a polynomial potential $V_\text{eff}(x)$, where $\mathbb{C}$ enters as a simple constant to get some density $\rho_\text{eff}(x|\mathbb{C})$. For example, for a single cut from $a$ to $b$ we would have as usual 
\beq
G(x)=\frac{i}{2\pi}\int_a^b \frac{\sqrt{(x-a)(x-b)}}{\sqrt{(y-a)(y-b)}}\frac{V_\text{eff}(y)}{x-y}dy \nn
\eeq
with the branchpoints $a,b$ fixed by imposing $G(x)\simeq 1/x$ at large $x$. At this point the solution depends on $\mathbb{C}$. Finally, we impose the self-consistency condition 
\beq
\mathbb{C}= \int \rho_\text{eff}(x|\mathbb{C}) \,x^q dx \nn
\eeq
to fix $\mathbb{C}$. For example, for a single cut solution in the $p=4$, $q=2$ model with 
\beq
Z_{4,2}= \int dX  \exp\Big(-\tfrac{N}{2} \tr X^{ 2}+
\tfrac{\alpha N^2}{{ 4}} \log (\Tr X^{ 4})
\Big)\label{Zabelian42}
\eeq
we would get at the end of the day
\beq
\rho(x)=\frac{\sqrt{a^2-x^2}}{6a^4\pi}(a^4+16x^2-4a^2(x^2-2))\quad \text{with}\quad \mathbb{C}=\frac{3a^4\alpha}{4(a^2-4)}, \quad a=2\sqrt{2-\sqrt{1-3\alpha}}  \la{rho42}
\eeq
Amusingly we would see that this solution has a phase transition and would cease to exist for~$\alpha>1/3$ at which point $\mathbb{C}$ -- which should be manifestly positive by definition -- would become complex. One could wonder if this solution dominates over the abelian case, perhaps for some range of $\alpha$. As we will see next, this is not the case and despite being a nice solution -- a nice new background -- it is always subleading in the path integral for $N$ large enough.  

\subsection{The Abelian Saddle Wins}
We found two saddles. Which is the dominant one? Since $x_1$ is very large in the Abelian saddle, the contribution to the effective action from the insertion of the huge operator would scale as 
\beq
\frac{\alpha N^2}{p} \log \Tr X^p
\simeq \frac{\alpha N^2}{p} \log \frac{x_*^p}{N}
\sim \alpha \Big(\frac{p}{q}-1\Big) N^2 \log N \nn
\eeq
which is enhanced by an extra $\log N$ compared to the more convencional $N^2$ scaling that the conventional saddle exhibits. Therefore the abelian saddle dominates and moments of light operators in the model (\ref{Zabelian}) would indeed be given by (\ref{xstarmoments}).

\begin{figure}
    \centering
    \includegraphics[width=\linewidth]{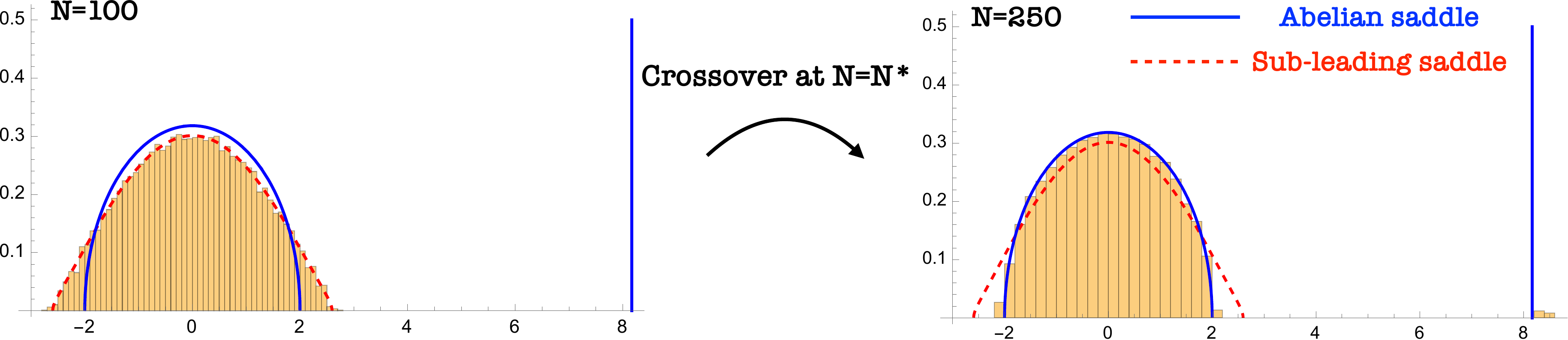}
    \caption{The left and right histograms are the results of Hybrid-Monte-Carlo simulations of the matrix model (\ref{Zabelian42}) for $N=100$ and $N=250$ respectively at $\alpha=0.3$. There is a transition at some critical $N_*$. At $N=100<N_*$ the background deforming saddle (\ref{rho42}) in red dominates while at $N=250>N_*$ we see nice agreement with the abelian saddle for which one eigenvalue decouples from the other ones -- which in turn are given by the standard vacuum semi-circle distribution.}
    \label{fig:enter-label}
\end{figure}

Sometimes we can have the more standard saddle dominate if $N$ is not large enough. This is precisely what we get for the example (\ref{Zabelian42}) for instance, see figure \ref{fig:enter-label}.

\subsection{Matrix Models and 
Huge Operators. When do we Abelianize?}
We found this new Abelianization when considering huge operators such as 
\beq
\mathcal{O}_p \equiv (\Tr X^{\color{red}p})^{\alpha N^2/{\color{red}p}} \label{hugeAb}
\eeq
What is special about such operators? Why did we never encounter this abelianization phenomena for all huge operator examples considered in previous sections? 

Potential operators will tend to localize the eigenvalues where the potential minima lie so they confine the eigenvalues there and thus do not allow for a single eigenvalue $x_1$ to fly away. Same for the classical source operators. As we saw in the previous sections, sources attract the eigenvalues towards the vicinity of the source location so provided we do not fine tune this sources abnormally with funny $N$ scalings the eigenvalues will stay close to the sources and again not escape. 

What about characters with $O(N^2)$ boxes? Why don't they Abelianize? The reason is that we have been implicitly considering characters associated to representations of generic Young-Tableaux with several rows. For example, in figure (\ref{YTexamples}) we have $N$ non-empty rows in each Young-Tableaux. Such operators with many rows vanish \textit{extremely} fast if we rescale all eigenvalues $x_2,\dots,x_N$ to zero. 
\beq
\chi_\mathcal{R}(x_1,\epsilon x_2,\dots,\epsilon x_N)= O(\epsilon^\mathcal{N}) \qquad , \qquad \mathcal{N}=\texttt{number of boxes in rows $2,3,\dots, N$}.  \label{epsilonN}
\eeq
Generically $\mathcal{N}$ scales as $N^2$. This is in sharp contradistinction with (\ref{hugeAb}) for which there is a finite~$\epsilon \to 0$ limit 
\beq
\lim_{\epsilon \to 0} O_p(x_1,\epsilon x_2,\dots,\epsilon x_N)= x_1^{\alpha N^2} \nn
\eeq
These two statements are of course compatible with each other because $O_p$ has finite overlap with very special symmetric characters with a single row,
\beq
O_p=1\times \chi_\texttt{single row $\mathcal{R}$ with $\alpha N^2$ boxes}+ \sum_\texttt{$\mathcal{R}$ with $\mathcal{N}>0$} c_\mathcal{R} \chi_\mathcal{R} \,. \nn
\eeq
In sum: a generic character vanishes very strongly but tuned operators like (\ref{hugeAb}) do not vanish at all. It is because of this that operators like (\ref{hugeAb}) favor this new type of Abelian saddles where one eigenvalue decouples from the others which we can therefore effectively set to zero. In other words, for abelian huge operators we can effectively drop all eigenvalues $x_2,\dots,x_N$ and think of (\ref{Zabelian}) as a one dimensional integral for a $U(1)$ theory with $x_1=X$. This we clearly can not do for general characters because of the huge supressions discussed above. 

The general recipe seems to be the following. Suppose we have an operator with $O(N^2)$ matrices. If the operator is finite when we set all but one eigenvalue to zero then the theory abelianizes into a $U(1)$ effective theory. It it vanishes very rapidly as in (\ref{epsilonN}),  it does not. 

This immediately opens the door to further interesting generalizations. If we have character operators with two huge rows, for instance, then we will get two non-trivial eigenvalues $x_1$ and $x_2$ and so on. This is very reminiscent of the tube limit in \cite{Abajian:2023jye,Kazakov:2024ald}. In AdS/CFT context, a single huge row is a giant graviton and two huge rows would thus correspond to two giant gravitons and so on.

\section{Final Speculative Thoughts: Going Beyond Hoppe}
One of the main reason to care about Yang-Mills like matrix models is to seek a dual description of quantum gravity in the universe. The Ishibashi-Kawai-Kitazawa-Tsuchiya (IKKT) matrix model \cite{Ishibashi:1996xs}, the Banks-Fischler-Shenker-Susskind (BFSS) matrix quantum mechanics \cite{Banks:1996vh} are notable supersymmetric proposals. Their mass deformed generalizations in the form of the Polarized IKKT matrix model \cite{Bonelli:2002mb,Hartnoll:2024csr,Komatsu:2024bop,Komatsu:2024ydh} or the Berenstein-Mandacena-Nastase mass deformed matrix quantum mechanics \cite{Berenstein:2002jq} are even richer proposals.  

The very simple Hoppe model \cite{hoppe:1989} studied in this paper, with two matrices $X$ and $Y$ is a nice toy model within this sort of models. Indeed, if we take the polarized IKKT model and drop all but two matrices we end up with the Hoppe model studied in this paper. We found universality in this model when the number of colours $N$ and the coupling $\lambda$ is large. We conjectured this universality to be related to the same sort of no-hair universality in Black Hole physics but this might be a stretch given the over simplicity of this model and of the probes considered herein. 

One direction we wish to pursue is to consider richer observables such as connected correlators in the Hoppe model. $\diamond$ Can we see that some of these correlators behave as sort of gravity massless modes while others have a massive behavior? This is of course non-trivial in a setup where space-time itself is emergent and thus distinguishing a powerlike from an exponential decay in a yet to be created space-time might be challenging. We should try. $\diamond$

Another direction we wish to pursue is to consider richer and richer models. Of course, the more intricate the model, the more likely it is that we will have to drop many of the analytic tools and rely more heavily on the Monte-Carlo numerics. This was in fact one of the motivations for honing these numerical tools in the Hoppe model in this work. 

\begin{figure}
    \centering
    \includegraphics[width=\linewidth,trim={1cm 5cm 0cm 0cm},clip]{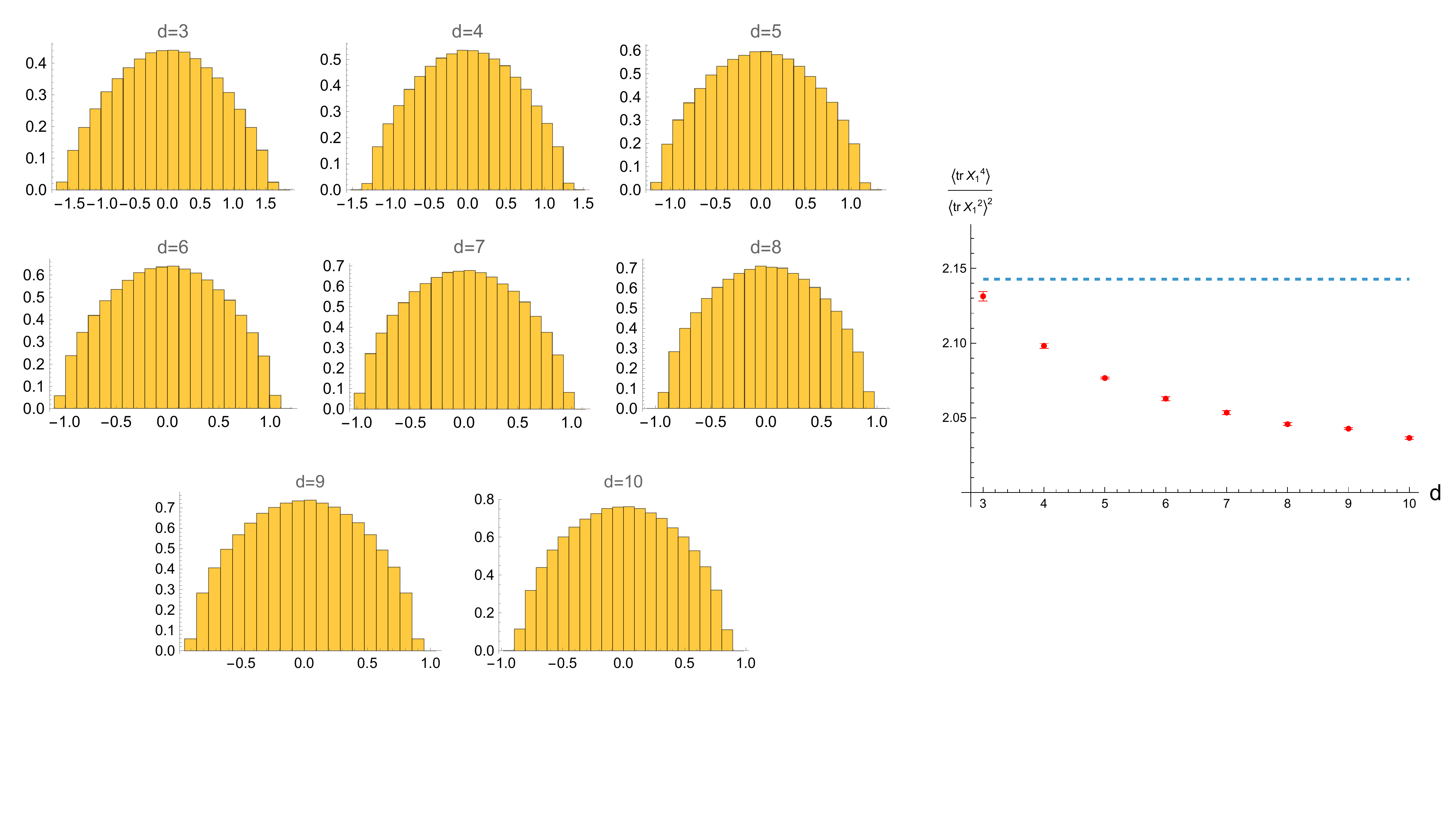}
    \caption{On the left, we have eigenvalue densities from HMC numerics for the massless $d>2$ Yang-Mills matrix models at $N=80$. Note that there are no couplings in this model. On the right, the red points are moment ratios \eqref{ratiod2} and the dashed blue line is what we expect for a parabolic density. Clearly, these densities are no longer parabolas and once we add sources, they get deformed in a non-universal manner.}
    \label{HMCdbiggerThan2}
\end{figure}

A most naive generalization of the Hoppe model would be the $d$ dimensional mass deformed Yang-Mills matrix model
\beq
S_d= \sum_{\mu=1}^d \tr (X_\mu)^2-\lambda \sum_{\mu,\nu=1}^d \tr [X_\mu,X_\nu]^2 \,. \nn
\eeq
For $d=2$ this is the Hoppe model. For $d>2$ the model behaves very differently. For one, two matrices $X_\mu$ with different $\mu$ no longer commute at strong coupling. The reason is a bit subtle. It turns out that the action $S_d^{\text{massless}}= -\lambda \sum_{\mu,\nu=1}^d \tr [X_\mu,X_\nu]^2$ defines a divergent matrix integral if $d=2$ but leads to a convergent matrix integral if $d>2$ \cite{Krauth:1998yu,Austing:2001bd,Austing:2001pk,Austing:2001ib}.\footnote{That the $d=2$ model diverges is simple to see. Take (\ref{mainEq1}) without any huge operator inserted for simplicity. Without the gaussian term, upon integrating out the $Y$ matrix we would now get $
\left\<{ \text{Tr}(X^n)} \right\> = \frac{\int dx_1\dots dx_N \, { x_1^n}  
}{\int dx_1\dots dx_N  1  }= \infty \, $ instead of  (\ref{mainEq}). Inserting a generic huge operator would not help. Polynomial huge operators such as characters would only make it worse. To see that the $d>2$ models converge is, on the contrary, considerably harder \cite{Austing:2001bd,Austing:2001pk}.} This has important consequences. It means that for very large $\lambda$ we can simply rescale the matrices by $X_\mu \to \frac{1}{\lambda^{1/4}} \hat X_{\mu}$ and effectively drop the mass term to obtain any moments such as
\beqa
\mathcal Z\times\< \tr [X_\mu, X_\nu]^2 \> &\simeq& \lambda^{-4/4} \int d \hat X_1 d \hat X_2 \dots d \hat X_d \,\tr [\hat X_\mu, \hat X_\nu]^2\, \exp{N\sum\limits_{\mu,\nu=1}^d\tr [\hat X_\mu,\hat X_\nu]^2}\nn\\
\mathcal Z\times\< \tr \{X_\mu, X_\nu\}^2 \> &\simeq& \lambda^{-4/4} \int d \hat X_1 d \hat X_2 \dots d \hat X_d \,\tr \{\hat X_\mu, \hat X_\nu\}^2\, \exp{N\sum\limits_{\mu,\nu=1}^d\tr [\hat X_\mu,\hat X_\nu]^2}\nn
\eeqa
In the right hand side of both lines we have a small number -- the $1/\lambda$ factor -- multiplied by an order one $\lambda$ independent finite matrix integral. Both lines are therefore equally small and their ratio is finite. In sharp contradistinction to the $d=2$ Hoppe model, we see that for $d>2$ it is \textit{not} true that the commutator is much smaller than the anti-commutator and we can no longer think of these matrices as commuting. Relatedly, even simple moments ratios such as 
\beq
\frac{\< \Tr (X_1)^4 \>}{\< \Tr (X_1)^2 \>^2} =\frac{\displaystyle \int \rho(x) x^4}{\Big(\displaystyle\int \rho(x) x^2\Big)^2 } \label{ratiod2}
\eeq
are not known at strong coupling in these higher dimensional Yang-Mills matrix models since even the eigenvalue density of a single matrix $X_1$ is not known. This is again in sharp contrast to the Hoppe model case where that density took a simple parabola form!, and for which the ratio (\ref{ratiod2}) simply evaluates to the universal ratio $15/7\simeq 2.14$ both for the vacuum as well as in the presence of a wide range of huge operators. In higher $d$ we computed the strong coupling vacuum density of $X_1$ eigenvalues and the corresponding moment ratios such as (\ref{ratiod2}) numerically using hybrid Monte-Carlo, see figure \ref{HMCdbiggerThan2}. By naked eye, we see that the result for $d=3$ or $d=4$ is close to a parabola but a quantitative analysis of the moment ratios in (\ref{ratiod2}) clearly indicates that the exact density is something else.

We see that the parabola \textit{simplicity} of the Hoppe model is not there for these higher d models. What about \textit{universality}? 
It is there albeit in a trivial way. Take a \textit{classical source}-type operator for instance so that $S \to S+\tr(J X)$. If the source $J$ does not scale with $\lambda$ it simply drops out just like the mass term once we rescale $X$ and we end up with the vacuum. For the source to \textit{do anything}, it should thus scale as $\lambda^{1/4}$ which is ``against the rules of the game". This is very different from the 2d case, where even an $O(1)$ source has interesting behavior, with phase transitions. The same conclusion would hold for other big operators such as characters or potential operators; the big operator might shift the center of mass distribution to a new locus -- e.g. if we insert an operator like $\det (X_\mu)^N$ the matrix eigenvalues will be repelled from the origin -- but otherwise, when expanding around that equilibrium point we will end up with the universal vacuum density of the massless Yang-Mills model. The punchline here is that higher dimensional purely bosonic massive Yang-Mills like matrix integrals will be even more universal than the Hoppe model but they will loose some of it's interesting structures such as the commuting matrices at strong coupling.

\begin{figure}[t]
    \centering
    \includegraphics[width=\linewidth]{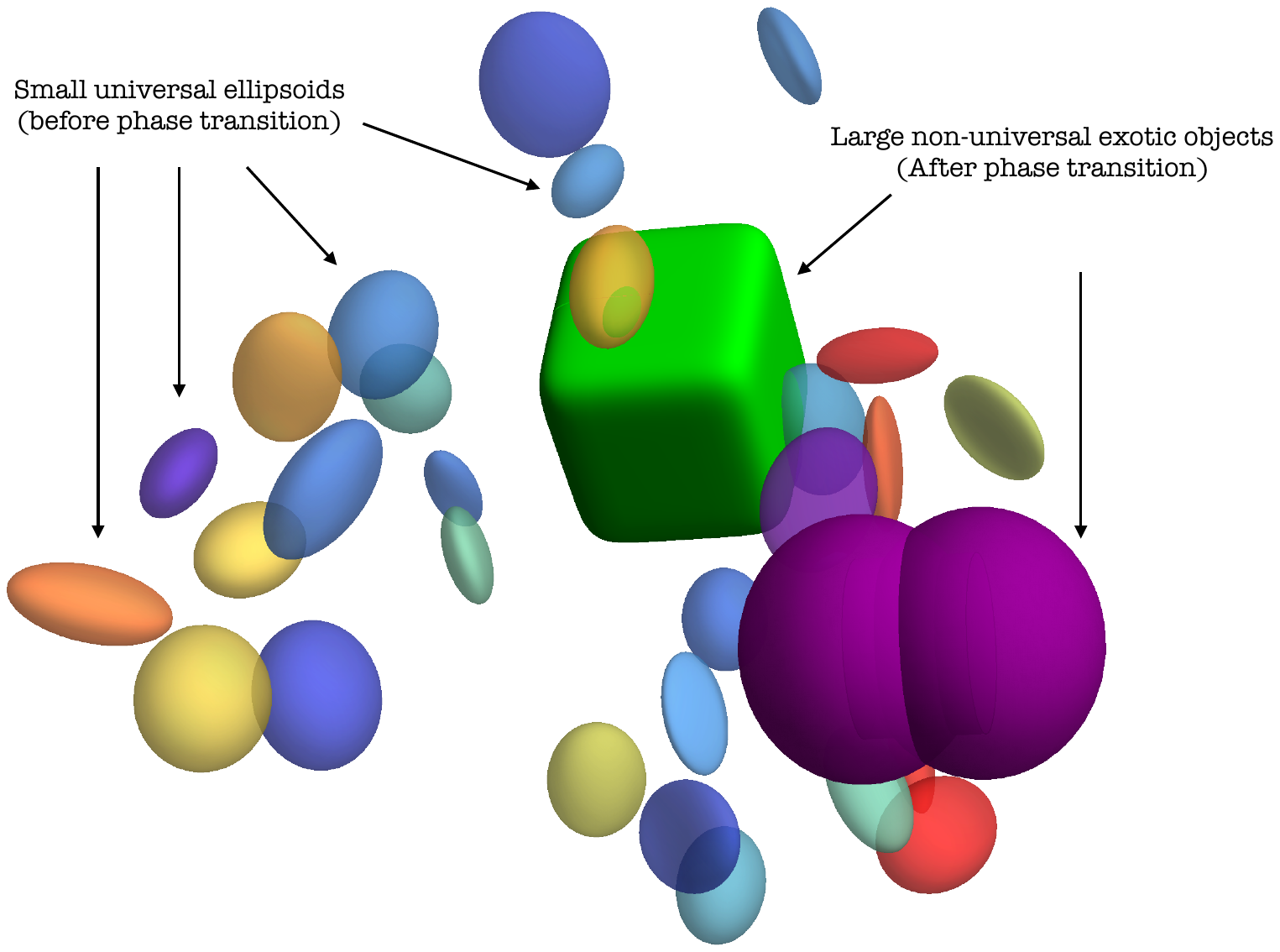}
    \vspace{-1cm}
    \caption{{\textbf{Matrix Model $\diamond$ Universes $\diamond$ at Strong Coupling.}} For the three dimensional model (\ref{XYZ}) we expect universal small ellipsoids of size $\lambda^{-1/6}$ before phase transitions and exotic objects depending on the details of the huge operator after phase transition. The two dimensional projection of this figure to two dimension -- with universal two dimensional ellipsoids as well as large non-universal 2D distributions like rings and so on -- is what we encountered over an over in this paper within the two-matrix Hoppe Model.}
    \label{theRealWorld}
\end{figure}

There are, however, some very nice higher dimensional bosonic models that retain that nice commutativity property. With three matrices, $X,Y,Z$,~\cite{Filev:2013pza} suggested
\beq
Z_{XYZ}= \int dX dY dZ \exp\Big( -N\tr\big(\tfrac{1}{2} X^2+\tfrac{1}{2} Y^2+\tfrac{1}{2} Z^2+\sqrt{2\lambda} Z[X,Y] \big) \Big)\,. \label{XYZ}
\eeq
as a beautiful example. This model has $SO(3)$ rotational symmetry. We could cast the last term as  $\epsilon_{abc} X^a X^b X^c$ to make that manifest. This type of interaction is what we commonly refer to as a Myers term \cite{Myers:1999ps}. Note that we can trivially integrate out $Z$ in (\ref{XYZ}) to get
\beq
Z_{XY}= \int dX dY \exp\Big( -N\tr\big(\tfrac{1}{2} X^2+\tfrac{1}{2} Y^2-\lambda [X,Y]^2 \big) \Big)\,. \label{XY}
\eeq
which is precisely the Hoppe model for which we know that at strong coupling we know that the matrices $X$ and $Y$ commute. Since we could have integrated out any of the three matrix, we conclude that $X$, $Y$ and $Z$ all commute with each other when $\lambda\gg 1$. In this limit, we can discuss the density of eigenvalues $\rho(x)$ of a single matrix as well as the joint distributions of eigenvalues of two matrices $\rho(x,y)$ as well as that of the three matrices, $\rho(x,y,z)$. The first two are the same as those in the Hoppe model while the last one is the unique rotationally symmetric uplift of the two dimensional hemispheres to three variables \cite{OConnor:2012vwc,Filev:2013pza} which turns out to be the simplest possible eigenvalue distribution: constant distribution inside a three dimensional ball!, 
\begin{equation}
\rho(x,y,z)=\frac{3}{4\pi} \frac{1}{L^3}\,,\qquad x^2+y^2+z^2 \le L^2 \, \qquad L=L_\text{vacuum}=\frac{(3\pi)^{\frac13}}{(2\lambda)^{\frac16}} \nn
\end{equation}
The lower point distributions are the hemisphere and the parabola encountered to exhaustion above, 
\begin{equation}
\rho(x,y)=\!\int \!dz\, \rho(x,y,z)=\frac{1}{L^2} \frac{3}{2\pi}\sqrt{1-\frac{x^2}{L^2}-\frac{y^2}{L^2}} \,, \qquad \rho(x)=\!\int\! dy\, \rho(x,y)= \frac{1}{L} \frac{3}{4}\Big(1-\frac{x^2}{L^2}\Big) \,. \nn
\end{equation}
This vacuum analysis was nicely done before in \cite{Filev:2013pza}; see also appendix \ref{uplifts}.

In the presence of huge operators, the round balls will be deformed into deformed ellipsoids with different axis $L_X$, $L_Y$ and $L_Z$.\footnote{More generally the three axis do not need to be aligned with these cartesian coordinates and would be a linear combination of the $X$, $Y$, $Z$ directions.} For example, if we insert operators that only depend on a single matrix $X$ then we established that the two dimensional distribution would be a deformed hemisphere in the universal phase, see (\ref{XYJointEllipse}). Since we should have rotational symmetry in the $Y,Z$ plane we conclude that this should lift to a three dimensional squashed ball. 
If the huge operator depends on three matrices, we expect again to obtain a squashed sphere without any particular rotational symmetry left but we can not proceed analytically in that case.

Can't we simply run some Hybrid Monte Carlo for (\ref{XYZ}) to verify this general squashed sphere behavior, study the universal and non-universal phases and so on? After all, we did run simulations for the various higher dimensional Yang-Mills models above with more matrices and more involved interactions... We can not! The reason is quite amusing: 

 
Note that $i[X,Y]$ 
is Hermitian, so the commutator $[X,Y]$ is anti-Hermitian and has purely imaginary eigenvalues. As a result, the interaction term $\mathrm{tr}\left(Z [X,Y]\right)$ is purely imaginary, making the full three-matrix model \eqref{XYZ}
non-unitary in the usual sense, and plagued by the infamous sign problem -- see \cite{Nagata:2021ugx} for a recent review. It instead belongs to the class of PT-symmetric matrix models, invariant under conjugation combined with flipping the sign of all matrices.
However, when one of the matrices is integrated out, as in reducing to the two-matrix model \eqref{XY}
the resulting theory has a real, bounded action and is perfectly unitary and free from any sign problem. Crucially, the sign of the commutator-square term ensures convergence for Hermitian matrices. If instead we tried to make the Myers term real by sending $\lambda \to -\lambda$, this would render the model divergent. In sum, the full three-matrix model exhibits mild non-unitarity due to its imaginary interaction, even though it reduces to a well-defined and unitary theory when one matrix is integrated out. When simulating the three matrix problem, however, we will encounter sign problem issues rendering numerics impractical with naive methods. (In a similar context, interesting numerical work on trying to face complex actions in Matrix models through clever sampling methods include \cite{Anagnostopoulos:2001yb, Ambjorn:2000dx,Ambjorn:2002pz}.) This is an amusing scenario where analytic computations remain tractable, while numerical simulations are obstructed. This suggests that the sign problem in this case may be less severe than those encountered in strongly correlated condensed matter systems \cite{Loh:1990zz}.\footnote{Of course, the simplest toy model of this sort of integral would be simply something like $I=\int_{-\infty}^{+\infty} dx e^{-x^2/2+i \alpha x}$. How to do such simple integrals and their higher dimensional extensions with Monte-Carlo? This simple integral is of course perfectly convergent but the $i\alpha x$ term is purely imaginary. We could combine positive and negative $x$ to get $I= \int_{-\infty}^{+\infty} dx e^{-x^2/2} \cos(\alpha x)$ but now the cosine has oscillating sign and we face the sign problem. We were told (by humans and AIs) that Complex Langevin might be a tool to learn.} 
Developing alternative numerical approaches, potentially inspired by machine learning, or complexified sampling strategies, could offer a way forward.


We can naturally construct many similar models involving more matrices. For example, consider for example $Z_{XYZW} = \int dX\, dY\, dZ\, dW e^{-N S_{XYZW}}$ with 
\begin{equation}
S_{XYZW}=\tr\left(\tfrac{1}{2} X^2 + \tfrac{1}{2} Y^2 + \tfrac{1}{2} Z^2 + \tfrac{1}{2} W^2 + \sqrt{2\lambda}\, Z[X,Y] + \sqrt{2\lambda}\, W[X,Y]+\lambda [X,Y]^2\right)  \,. \nn
\end{equation}
Integrating out $W$ cancels the last term and leads to the previous $XYZ$ model and therefore $X,Y,Z$ all commute; similarly, integrating out $Z$ would lead us to conclude that $X,Y,W$ all commute. If these were finite matrices with non-degenerate eigenvalues this would mean that $[Z,W]$ should vanish as well. To check if it really does 
we can compute $\<\tr[W,Z]^2\>$ explicitly integrating out~$Z$ and~$W$ to reduce this quantity to a bunch of moments in the Hoppe model. We get, at large $N$ and strong coupling,
  \beq
 \<\tr [Z,W]^2\>_{XYZW}=-2 - 8 \lambda \underbrace{\<\tr [X,Y]^2\>_\text{Hoppe} }_{=-\frac{1}{2\lambda}}=-2+4\neq 0 \,, \nn
 \eeq
and thus they unfortunately do not commute\footnote{Note that the measure is not positive here, which is why we get $\langle\tr [Z,W]^2\rangle>0$. If the measure were positive, we would have a small commutator since $(-\langle \Tr[Z,W]^2\rangle) \leq 2 \langle\Tr Z^2\Tr W^2\rangle\sim \lambda^{-2/3}$ for positive measures.} -- commutativity is not transitive at large N! Perhaps there is a simple deformation of this model where they do. 
Despite the apparent simplicity of such four-matrix model, we cannot straightforwardly explore its full structure using Hybrid Monte Carlo due to the same sign problem discussed earlier. It would be fascinating to chart the broader landscape of bosonic matrix models that exhibit commuting behavior at strong coupling. We believe that simple structures like Myers and Yang-Mills terms can serve as building blocks in such an exploration. Would be great if  we could find some simple commuting bosonic matrix model without funny imaginary couplings and associated sign problem. We leave a systematic exploration of this possibility to future work.


Going back to the richer matrix models like IKKT and polarized IKKT, one can integrate out the fermionic degrees of freedom which yields an insertion of a Pfaffian. Typically, this term is not positive definite, leading once again to the sign problem. In the $D=4$ IKKT model the Pfaffian is real and positive~\cite{Krauth:1998xh,Ambjorn:2000bf}. One can even add to it a supersymmetric mass term~\cite{Adrien} akin to the polarized IKKT model. However, large $N$ Monte Carlo simulations are still non-trivial and one encounters challenges due to the high computational cost and potential numerical instability of evaluating derivatives. In this context, normalizing flows~\cite{Abbott:2023thq} offer a promising alternative. By training an invertible neural network to transform a simple base distribution into one approximating the target distribution defined by the matrix model action, one can effectively circumvent the need for explicit gradient computation during sampling. Recent work has demonstrated the successful application of flow-based sampling to study the dynamics of confining strings~\cite{Caselle:2023mvh,Caselle:2024ent}, suggesting potential for broader use in matrix models with complicated interactions.
Another potentially useful tool to overcome the computational cost of Monte-Carlo for the $D=4$ IKKT model is the so called Matrix Model Bootstrap \cite{AndersonKruczenski2017,LinBootstrapsToStrings2020,KazakovZhengAnalyticNumericalBootstrap2022,LinD0BraneQuantumMechanicsBounds2023,LiZhouBootstrappingAbelianLatticeGaugeTheories2024}. There, we directly work at large $N$ and impose positivity and loop equations to obtain bounds on correlation functions. In particular, in \cite{KazakovZhengAnalyticNumericalBootstrap2022}, these methods were used to solve the Hoppe model with a quartic potential at finite $\lambda$. Would be fascinating if we could rederive and extend some of the results presented here with such tools. 
In \cite{LinD0BraneQuantumMechanicsBounds2023,Lin:2024vvg} the bootstrap was used to bound interesting quantities in the BFSS matrix quantum mechanics. Closely related to our setup, \cite{Hoppe:1999xg} might be a great starting point for such explorations. Also in this vein, it would be great to solve the loop equations directly in the presence of sources using topological recursion as in \cite{Vescovi:2024fwt}.

This work is complementary to \cite{Kazakov:2024ald}, studied by two of us together with Vladimir Kazakov. \textit{Here}, character operators of many different shapes lead to the same parabolic eigenvalue distributions. \textit{There}, different characters lead to different eigenvalues distributions governed by rich fluid dynamics with different initial conditions. \textit{There}, the non-universality was expected since we were studying interactions of horizonless geometries -- so called LLM geometries \cite{LLM} -- that do have a ton of details. Indeed, every coloring of the plane into black and white regions leads to a unique bulk geometry and one can map different characters to different colorings (see for instance fig. 8 in \cite{LLM}). \textit{Here} we are after black hole like universality where no-hair theorems ought to lead to important emergent universality. \textit{There} we were studying $\mathcal{N}=4$ SYM for which there is a clear gravity dual while here we are studying a much more trivial toy model. \textit{There}, computations were possible because of the huge amount of supersymmetry; \textit{here} we have no supersymmetry but computations are possible because we are dealing with a simple matrix model. Would be nice to start from these two examples as end-points and meet somewhere in the middle by trying to consider less supersymmetric observables in $\mathcal{N}=4$ SYM or by trying to add more dimensions and/or supersymmetry to the perspective put forward here. In the middle we will need computers; we should be describing real quantum black hole physics after all. 


\section*{Acknowledgments} 
We thank K.~Budzik, D.~Gaiotto, A.~Holguin, V.~Kazakov, S.~Komatsu, A. Martina, J.~Penedones, A.~Vuignier, X.~Zhao for numerous enlightening discussions. We thank V.~Kazakov and J.~Penedones for important suggestions and comments on the draft. HM and PV thank EPFL for hospitality during part of this work. 
Research at Perimeter Institute is supported in part by 
the Government of Canada through the Department of Innovation, Science, and Economic Development Canada and by the Province of Ontario through the Ministry of Colleges
and Universities. 
PV is supported in part by Discovery Grants from 
the Natural Sciences and Engineering Research Council of Canada, 
and by the Simons Foundation through the ``Nonperturbative Bootstrap'' collaboration (488661). 
A.G. is supported by a Royal Society University Research Fellowship, URF/R1/241371.
This work was additionally supported by   
FAPESP Foundation through the grants 2016/01343-7, 2017/03303-1, 2020/16337-8. 

\appendix
\section{Large $N$ HCIZ integral as Hopf flows}\label{fluidAppendix}
In this appendix, we review results about the large $N$ limit of the Harish-Chandra-Itzykson-Zuber (HCIZ) integral first studied in \cite{Matytsin_1994}. Let us recall the definition of the HCIZ integral,
\begin{align}
    I(a,b)&\equiv \int dU\, e^{N\tr AU^\dagger BU} \label{hcizDefn}\\
    &= \frac{\det_{ij}e^{N a_i b_j}}{\Delta(a)\Delta(b)} \times \underbrace{\frac{G(N+1)}{N^{N(N-1)/2}}}_{\equiv \mathfrak{N}}
    \label{hcizFiniteN}
\end{align}
where $U$ is an $N\times N$ unitary matrix, $A$ and $B$ are Hermitian matrices with eigenvalues $\{a_i\}$ and $\{b_i\}$ and $G(N)$ is the Barnes G-function. The numerical constant $\mathfrak{N}$ does not affect any saddle point equations and we will ignore it. The second equation above is an exact finite $N$ result that can be derived in several ways including localization, heat kernel method and Dyson's Brownian motion. 

In \cite{Matytsin_1994}, Matytsin showed that at large $N$ the HCIZ integral has a beautiful interpretation in terms of a one-dimensional frictionless fluid flow obeying the Hopf equation. At large $N$ the eigenvalues condense into the continuum flow which extends from $t=0$ to $t=1$ where the following boundary conditions are imposed on the fluid's density $\rho(x,t)$,
\begin{equation}
    \rho(x,t=0) = \sigma_1(x) \ ,\qquad \rho(x,t=1) = \sigma_2(x)\label{HCIZfluidFlowBC}
\end{equation}
where $\sigma_1(a)$ and $\sigma_2(b)$ are the eigenvalue densities of $A$ and $B$ respectively. Along the way, the flow obeys the usual inviscid Navier-Stokes equations in 1+1d\footnote{Note that we could also package these equations into a single complex PDE by defining $f(x,t)=v(x,t)+i\pi \rho(x,t)$ which obeys the Riemann-Hopf equation $\partial_tf+f\partial_xf=0$ for the equation of state \eqref{equationOfState}.},
\begin{equation}
\begin{aligned}
    \frac{\partial \rho}{\partial t} + \frac{\partial(\rho v)}{\partial x}&=0\\
    \frac{\partial v}{\partial t} + v \frac{\partial v}{\partial x}+\frac{1}{\rho}\frac{\partial P}{\partial x}&=0
\end{aligned}\label{HCIZfluidFlowEquations}
\end{equation}
where $\rho$, $v$ and $P$ are the density, velocity and pressure of the fluid. For the HCIZ flow, the fluid has the following equation of state
\begin{equation}
    P = -\frac{\pi^2}3 \rho\label{equationOfState}
\end{equation}
which tells us that the fluid is somewhat exotic with a negative pressure. Once we have solved the PDEs \eqref{HCIZfluidFlowEquations} with the boundary conditions \eqref{HCIZfluidFlowBC}, we can compute the asymptotics of the HCIZ integral as,
\begin{equation}
    \frac1{N^2}\log I(a,b) = S_{\texttt{fluid}}[\rho,v] + S_{\texttt{boundary}}[\sigma_1,\sigma_2] + O\left(\frac1{N^2}\right)\label{largeNHCIZFluidBdy}
\end{equation}
where $S_{\texttt{fluid}}$ is given as usual by difference of kinetic and potential energies
\begin{equation}
    S_{\texttt{fluid}}[\rho,v] = \int_0^1 dt\int dx\, \rho(x,t)\left(v(x,t)^2 + \frac{\pi^2}3\rho(x,t)\right)\label{sfluidOnly}
\end{equation}
The second term in \eqref{largeNHCIZFluidBdy} is a boundary term which only depends on the end-point densities $\sigma_1(a)$ and $\sigma_2(b)$
\begin{equation}
    S_{\texttt{boundary}}[\sigma_1,\sigma_2] = \frac12 \sum_{i=1}^2\left[ \int dx\, x^2 \sigma_i(x) -\int dx\,dy\, \log|x-y|\sigma_i(x)\sigma_i(y)\right]-\frac34 \nn
\end{equation}

A nice feature of the equations (\ref{HCIZfluidFlowEquations},\ref{equationOfState}) is that they are \textit{integrable}. As shown in \cite{Matytsin_1994} and extended in \cite{Kazakov:2024ald}, we can write down infinitely many conserved charges for the fluids. To do so, we first define two functions,
\begin{equation}
    G_+(x)=x+v(x,0)+i\pi\rho(x,0) \ , \qquad G_-(x) = x-v(x,1)-i\pi\rho(x,1)\label{GpmDefns}
\end{equation}
in terms of which, the conservation laws read
\begin{equation}
    Q_{n,m} = \frac1{m+1}\oint \frac{dx}{2\pi i} x^n G_+(x)^{m+1} = \frac{-1}{n+1}\oint \frac{dx}{2\pi i}x^m G_-(x)^{n+1} \nn
\end{equation}
In particular, note that $Q_{n,0}$ are the moments of $\sigma_1$ and $Q_{0,m}$ are the moments of $\sigma_2$.

Another interesting fact about this flow is that the PDEs can be converted into the following functional equations,
\begin{equation}
\begin{aligned}
    G_+\left(G_-(x)\right) &= x \nn\\
    \text{Im } G_+(x) &= \pi\sigma_1(x) \nn\\
    \text{Im }G_-(x) &= -\pi\sigma_2(x) \nn
\end{aligned}
\end{equation}
That is, the $G_+$ and $G_-$ defined above are inverses of each other. Let us now use this technology to find the universal parabola parameters at strong coupling.

\subsection*{Parabolas for Classical Sources}

\begin{figure}[t]
    \centering
    \includegraphics[width=0.7\linewidth]{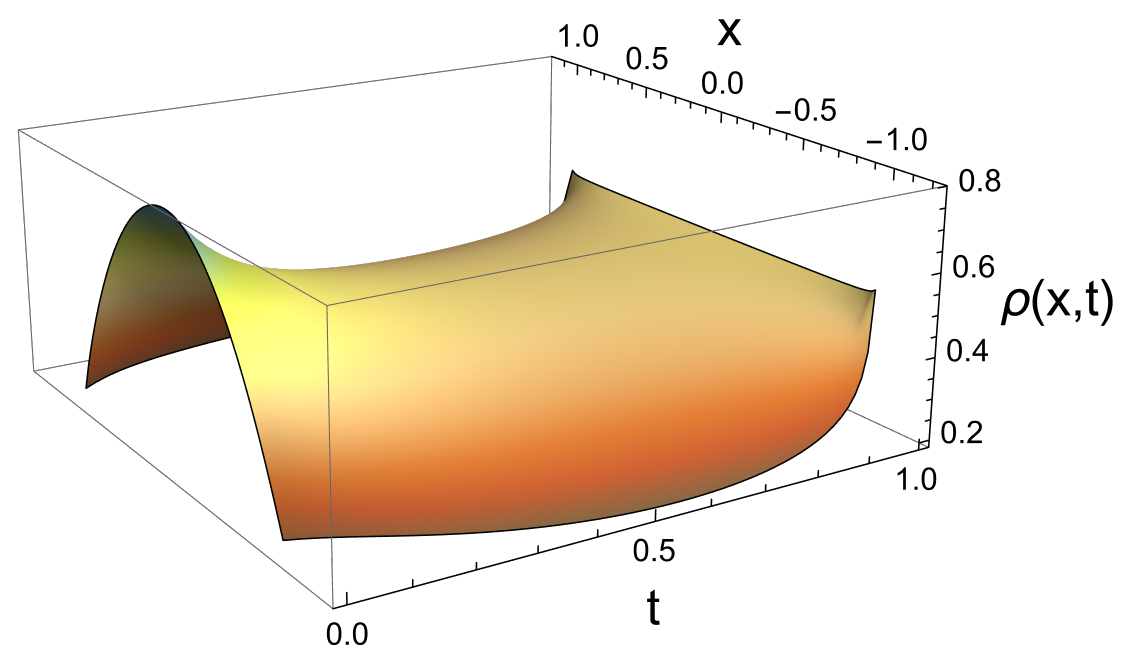}
    \caption{HCIZ fluid flow for the uniform line source with $\alpha=2$ and $\lambda=100$. At $t=0$, the density is parabolic (only approximately, because $\lambda=100\neq\infty$) and at $t=1$, we have a constant density of $\tfrac12$ on the interval $(-1,1)$.}
    \label{flowDensityFig}
\end{figure}

For convenience, we reproduce here the Hoppe model with a classical source \eqref{Iintegral},
\beq
    \int \prod_i dx_i\, \Delta_\lambda(x)^2 e^{-\frac N2 \sum_i x_i^2}\, I(x,j)\nn
\eeq
Now, we have single HCIZ flow from $I(x,j)$ that goes from $\{x_i\}$ at $t=0$ to $\{j_i\}$ at $t=1$. We also have the saddle point equations for the $x_i$ integrals which take the form \footnote{In computing the SPEs we need to evaluate $\frac{\partial S_{\texttt{fluid}}}{\partial \rho(x,0)} = v(x,0)$ which comes from the boundary of the $t$-integral in \eqref{sfluidOnly}. Similarly, we have $\frac{\partial S_{\texttt{fluid}}}{\partial \rho(x,1)}=-v(x,1)$.}
\beq
    -x + \underbrace{2\fint dy\, \frac{\sigma_1(y)}{(x-y)(1+2\lambda(x-y)^2)}}_{\texttt{from $\Delta_\lambda(x)^2$}} + \underbrace{\left(x - \fint dy\, \frac{\sigma_1(y)}{x-y}\right)}_{\texttt{from $S_{\texttt{boundary}}$}} + \underbrace{v(x,0)}_{\texttt{from $S_{\texttt{fluid}}$}}=0\label{speXVelSource}
\eeq
Therefore, knowing $\sigma_1(x)$ we can compute $v(x,0)$ and with that we can solve the flow equations. Of course, we don't know the density $\sigma_1(x)$ but for some special cases like the homogenous source of section \ref{uniformLineSection}, we can find the density approximately by discretizing the SPEs \eqref{SPalpha} and solving them numerically. Then, we can find the initial velocity $v(x,0)$ and do the flow evolution\footnote{In practice, we solve the rational Calogero-Moser equations which are a discretization of the Hopf flow. See \cite{Kazakov:2024ald} for details} to obtain $\rho(x,t)$, which is shown in figure \ref{flowDensityFig}.

For more general sources, we don't have simple SPEs for the eigenvalues. So instead let us make an ansatz for the density $\sigma_1(x)$ -- a natural ansatz is a parabola centered at $x_0$ and with size $L$. Plugging in this density into \eqref{speXVelSource} and going to strong coupling we get
\beq
\begin{aligned}
    v(x, 0) - \fint dy\, \frac{\sigma_1(y)}{x-y} &= -2\fint dy\, \frac{\s_1(y)}{(x-y)(1+2\lambda(x-y)^2)}\nn \\
    &= -\frac{L_{\text{vacuum}}^3}{L^3}(x-x_0) + \ldots \nn
\end{aligned}\label{kernelStrongCouplingSourceSimp}
\eeq
Plugging this into the definition of $G_+(x)$ \eqref{GpmDefns}, we get 
\beq
    G_+(x) = x + R_1(x) -\frac{L_{\text{vacuum}}^3}{L^3}(x-x_0) + \ldots\label{GplusSourceExpansion}
\eeq
where $R_1(x)=\int dy\, \frac{\sigma_1(y)}{x-y}$ is the resolvent of $x$ obtained by combining the principal part integral with the density $i\pi \sigma_1(x)$ in \eqref{GpmDefns}. Now, we will use the conserved charges to fix the parabola parameters. Recall that conservation of $Q_{0,n}$ implies
\beq
    \Tr J^n = \frac1{n+1} \oint \frac{dx}{2\pi i}\, G_+(x)^{n+1}  \nn
\eeq
where we used the fact that the contour integral over $G_-(x)$ simply picks up the moments of $J$. We can evaluate the above contour integrals by picking up the residues at infinity. Note that the resolvent at infinity looks like $R_1(x) = \frac1x + \frac{\langle\Tr X\rangle}{x^2}+\frac{\langle\Tr X^2\rangle}{x^3}+\ldots$. Now, for $n=1$, we have
\beq
    \Tr J = \langle \Tr X\rangle = x_0 \nn
\eeq
which fixes the center of the parabola to be at the center of mass of the source $J$. Now let's look at the $n=2$ charge
\beq
    \Tr J^2 = \left[1 - \frac{L_{\text{vacuum}}^3}{L^3} + (\text{Tr}\, J)^2\right] + \frac{L^2}5\left(\frac{L_{\text{vacuum}}^3}{L^3}-1\right)^2 + \ldots \nn
\eeq
where the term outside the square paranthesis is subleading because $L\ll 1$. Therefore, we recover the expression \eqref{parabolaForJX} from the main text. The higher charges will involve subleading terms in \eqref{GplusSourceExpansion} which are non-universal. We can try to learn what these corrections are using the charges -- it would be nice to compare these with the strong coupling expansions found in \cite{Filev:2013pza} and \cite{Berenstein:2008eg}.
\subsection*{Parabolas for Characters}
For characters $\chi_R(X)$, we need to analyze the following integral,
\beq
    \int \prod_i dx_i\, \Delta_\lambda(x)^2 e^{-\frac N2 \sum_i x_i^2}\, I(\log x,h) \frac{\Delta(\log x)}{\Delta(x)}\nn
\eeq
where we dropped some factors from \eqref{characterAsHCIZ} which do not affect saddle points in $x$. This integral can also be described by a fluids but now, the flow goes from $\{\log x_i\}$ at $t=0$ to $\{h_i\}$ at $t=1$. The SPE for $x$ is
\begin{multline}
    -x + \underbrace{2\fint dy\, \frac{\sigma_1(y)}{(x-y)(1+2\lambda(x-y)^2)}}_{\texttt{from $\Delta_\lambda(x)^2$}} + \underbrace{\frac1x\left(\log x - \fint dy\, \frac{\sigma_1(y)}{\log x-\log y}\right)}_{\texttt{from $S_{\texttt{boundary}}$}} \\+ \underbrace{\frac1x\,v(\log x,0)}_{\texttt{from $S_{\texttt{fluid}}$}} + \underbrace{\fint dy\, \sigma_1(y)\left(\frac1{\log x-\log y} - \frac1{x-y}\right)}_{\texttt{from $\frac{\Delta(\log x)}{\Delta(x)}$}} =0\label{speXVelChar}
\end{multline}
Now, we once again make the ansatz that $\sigma_1(x)$ is a parabola with center at $x_0$ and size $L$. The conserved charges we are interested in are,
\beq
    \underbrace{-\oint \frac{dx}{2\pi i}\, x^n G_-(x)}_{=\langle h^n\rangle} = \frac1{n+1}\oint \frac{d\log x}{2\pi i}\, G_+(\log x)^{n+1} \nn
\eeq
where $G_+(\log x)$ is given by
\beqa
    G_+(\log x) &=& \log x + v(\log x,0) + i\pi\, x\s_1(x)\nn\\
    &=& x\(x + \int dy\, \frac{\s_1(y)}{x-y} - 2\fint dy\, \frac{\s_1(y)}{(x-y)(1+2\lambda (x-y)^2)}\)\nn\\
    &=& x\(x + R_1(x) - \frac{L_{\text{vacuum}^3}}{L^3} (x-x_0)\) + \ldots \nn
\eeqa
In the second equality above, we plugged in the initial velocity $v(\log x,0)$ that follows from the SPE for $x$. Just like in the classical source case, the principal value integral combines with the density $\sigma_1(x)$ to give the full resolvent $R_1=\int dy\,\frac{\s_1(x)}{x-y} $ and we get the third equation at strong coupling. Proceeding in complete analogy with the case of classical source, the $n=1$ and $n=2$ charges fix the parabola parameters shown in \eqref{parabolaParamsCharacter}.

\section{Moments and Joint Density Distributions}
\subsection{Uplifts} \label{uplifts}
Given a three dimensional distribution $\rho(x,y,z)$ we can obtain the two dimensional projection $\rho(x,y)$ by integrating over $z$ and the one dimensional projection $\rho(x)$ by integrating over $y$ as well. Going the other way to obtain $\rho(x,y)$ given $\rho(x)$ or $\rho(x,y,z)$ given $\rho(x,y)$ can be easily done if we have spherical symmetry. In that case, the so-called inverse Abel transform relates  
\beq
\rho(x,y,z)=f_{3D}(\sqrt{x^2+y^2+z^2}) \,, \qquad \rho(x,y)=f_{2D}(\sqrt{x^2+y^2})\,, \qquad \rho(x)=f_{1D}(x)  \nonumber
\eeq
as\footnote{Here we assumed that the density have a support from the origin all the way to a maximum radius~$L$ otherwise one should simply replace $L$ by $\infty$ in these formulae.}
\beq
f_{3D}(r)=- \frac{1}{\pi}\int_r^L \frac{ dw}{ \sqrt{w^2-r^2}}\,f_{2D}'(w) \,, \qquad f_{2D}(r)=- \frac{1}{\pi}\int_r^L \frac{dw}{ \sqrt{w^2-r^2}}\, f_{1D}'(w)  \nonumber
\eeq
For our examples we would get 
\beq
f_{1D}=\texttt{parabola}=\frac{L^2-r^2}{4L^3/3} \,\,\, \Rightarrow \,\,\, f_{2D}=\texttt{hemisphere}=\frac{\sqrt{L^2-r^2}}{2L^3 \pi/3}  \,\,\, \Rightarrow \,\,\,  f_{3D}=\texttt{uniform}=\frac{1}{4\pi L^3/3}  \,. \nonumber 
\eeq
If we keep playing this game going to higher dimensions we see that for $4D$ the density  diverges as a square root at the shell $r=L$; we can not go to five or higher dimensions. 

\subsection{$1D \to 2D$} \label{WickAppendix}

\begin{figure}[t]
    \centering
    \includegraphics[width=\linewidth,trim={1cm 5cm 0cm 0cm},clip]{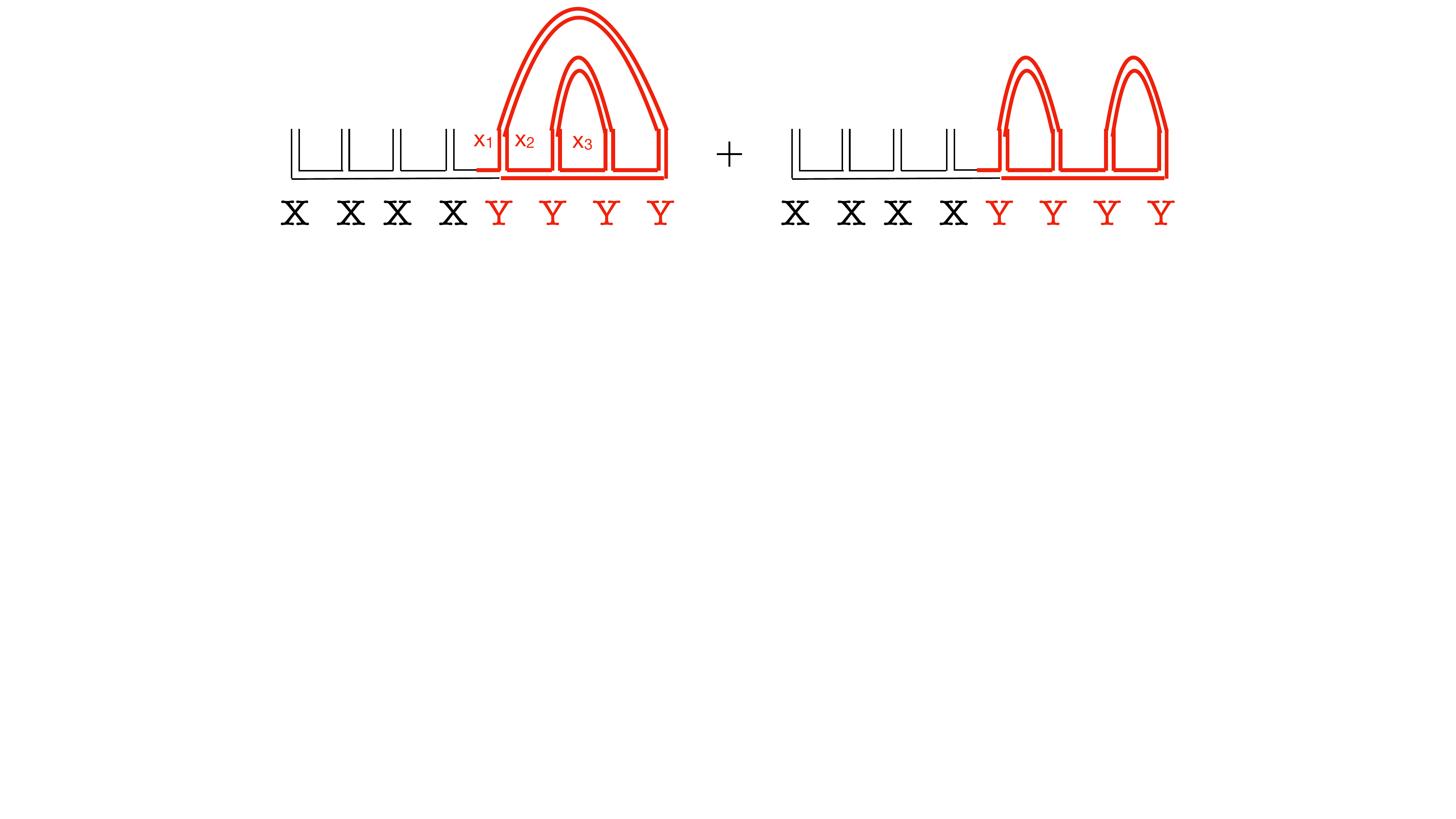}\vspace{-5cm}
    \caption{Wick contractions relevant for the evaluation of $\< \tr(X^4 Y^4) \> $ in the planar limit.}
    \label{exampleYs}
\end{figure}
Here we give a few details on a more general Wick contraction of more $Y$'s. The propagator for $Y$'s is
\beq
    \wick{\c Y_{ij} \c Y_{kl}} = \frac{\delta_{il}\delta_{kj}}{N}  \frac1{1+2\lambda(x_i-x_j)^2} \nn
\eeq
Note that in the t'hooft double lines, each line has an associated $x_i$. Let us consider as an example the computation of $\< \Tr(XXYXYXYY) \> $. The first thing to realize is that since $X$'s and $Y$'s commute at strong coupling this moment gives the same as
\beq
\< \Tr(XXXXYYYY) \>  \nn
\eeq
There are of course three Wick contractions we can do between the $Y$'s but in the planar limit only the $2$ in figure \ref{exampleYs} survive. This $2$ is the second Catalan number $C_2=2$; If we had a correlator with~$6$~$Y$'s we would have three propagators and $C_3=5$ diagrams contributing in the planar limit and so on. In general, for $2k$ $Y$'s, we need to count the number of rainbow graphs which is equal to the Catalan number $C_{k}$. Let us now observe that these various diagrams evaluate to the same thing using our example again as illustration. Using the propagator above, we have the two contribitions in the figure so that 
\beq
\!\!\!\! \< \Tr(X^4 Y^4) \> =\frac{1}{N^3} \sum_{i,j,k} x_i^4\, \frac1{1+2\lambda x_{ij}^2} \, \frac1{1+2\lambda x_{jk}^2}+\frac{1}{N^3} \sum_{i,j,k} x_i^4\, \frac1{1+2\lambda x_{ij}^2} \, \frac1{1+2\lambda x_{jk}^2} \nn
\eeq
with $x_{ij}=x_i-x_j$.
Next we convert each sum into an integral and use that in the large $\lambda$ limit the various Lorentzians become delta functions as~\footnote{One might worry that we have $\lambda\gg1$ but $x\ll1$. But this is not a problem because the support scales as $L= O(\lambda^{-1/6})$ and so $\lambda L^2\gg 1$.}
\beq
\frac{1}{1+2 \lambda x^2}= \frac{\pi}{\sqrt{2\lambda}} \delta(x) \nn 
\eeq
to get 
\beqa
 \<\Tr(X^4 Y^4) \> &=&\left(\frac{\pi}{\sqrt{2\lambda}} \right)^2 \int dx \int dy \int dz \rho(y) \rho(y) \rho(z) x^4 \delta(x-y)\delta(y-z) \nn \\
 &&+
\left(\frac{\pi}{\sqrt{2\lambda}} \right)^2 \int dx \int dy \int dz \rho(y) \rho(y) \rho(z) x^4 \delta(x-y)\delta(x-z) \,. \nn
\eeqa
The delta functions get rid of all but one integral so that in the end we get 
\beqa
 \< \Tr(X^{\color{red}4} Y^{\color{blue} 4}) \> =2\times \left(\frac{\pi}{\sqrt{2\lambda}} \right)^{{\color{blue} 4}/2} \int dx \,x^{\color{red}4} \rho(x)^{1+{\color{blue}4}/2} \,. \nn
\eeqa
which is of course a special case of 
\beqa
 \<\Tr(X^{\color{red}n} Y^{\color{blue} m}) \> =C_{{\color{blue} m}/2}\times \left(\frac{\pi}{\sqrt{2\lambda}} \right)^{{\color{blue} m}/2} \int dx\, x^{\color{red}n} \rho(x)^{1+{\color{blue}m}/2} \,. \nn
\eeqa
presented in the main text. The general case goes through in exactly the same way -- we have $C_{m/2}$ rainbow graphs all of which evaluate to the same integral shown above at strong coupling. Knowing all these moments was what we needed in the text to uplift the $X$ distribution $\rho(x)$ to the joint distributions $\rho(x,y)$, see (\ref{2drhoxy}).

\section{Hybrid Monte Carlo Algorithm Implementation}
\label{app:HMC}

This appendix outlines the Hybrid Monte Carlo (HMC) algorithm used in our simulations. We describe the general algorithmic structure, independent of the specific model under consideration.
The algorithm described below is based on~\cite{Jha:2021exo}, adapted to our specific setting. 

The HMC algorithm is a Markov Chain Monte Carlo (MCMC) method that combines molecular dynamics with a Metropolis accept/reject step. It enables efficient sampling of configuration space by reducing autocorrelations between successive configurations. The algorithm introduces a fictitious time evolution governed by a Hamiltonian, ensuring both ergodicity and detailed balance.

\paragraph{Dynamical Variables and Hamiltonian}

The dynamical variables consist of a set of $N_\text{mat}$ traceless Hermitian matrices $X_i$ of size $N \times N$. Each $X_i$ is paired with a conjugate momentum matrix $P_i$ of the same type. The fictitious Hamiltonian governing the dynamics is given by
\begin{equation}
    H = \sum_{i=1}^{N_\text{mat}} \frac{1}{2} \operatorname{tr}(P_i^2) + V(X), \nn
\end{equation}
where $V(X)$ is the potential energy. In most of our examples, it includes the Hoppe potential (i.e., the square of the commutator) and a model-dependent term such as $\log \mathcal{O}_\text{huge}$. The initialization of $X_i$ depends on the potential $V(X)$; when chosen carefully, it can lead to faster thermalization. The matrices $P_i$ are random Hermitian matrices whose elements are sampled from a Gaussian distribution with unit variance.

\paragraph{Leapfrog Integration}

At each Monte Carlo step $n$, a new configuration $X(n+1)$ is proposed by evolving the previous configuration $X(n)$ for a fictitious time 
$T = n_\text{steps} \times dt,$
using the Hamiltonian above with randomly sampled momenta.

The phase-space evolution is approximated using the leapfrog integrator, which is symplectic and time-reversible. The integration proceeds as follows:

\begin{enumerate}
    \item Sample each momentum $P_i$ from a Gaussian distribution.

    \item Perform a half-step update of the coordinates:
$X(n) \rightarrow X(n) + \frac{dt}{2} \, P(n).$

    \item Repeat the following for $n_{\text{steps}}$ steps:
    \begin{enumerate}
        \item Compute the force:
$F(n) = -\frac{\partial V}{\partial X}\biggr|_{X = X(n)}.$

        \item Update the momenta:
$P(n) \rightarrow P(n) - dt \, F(n).$

        \item Update the coordinates:
$X(n) \rightarrow X(n) + dt \, P(n).$
    \end{enumerate}

    \item Perform a final half-step update of the coordinates:
$X(n) \rightarrow X(n) + \frac{dt}{2} \, P(n).$
\end{enumerate}

The step size $dt$ controls how accurately the dynamics follow the Hamiltonian trajectory, while $n_\text{steps}$ determines the total integration time. In our simulations, we typically used $dt = 0.5 \times 10^{-4}$ and $n_\text{steps} \sim 300$.

\paragraph{Metropolis Acceptance Step}

After the integration, the Hamiltonian is re-evaluated. The proposed configuration $(X_i', P_i')$ is accepted with probability
$\mathcal{P}_\text{accept} = \min\left(1, e^{H_\text{initial} - H_\text{final}} \right).$
If the proposal is rejected, the original configuration is retained.

\paragraph{Observables and Monitoring}

After each accepted step, observables -- such as the eigenvalue distributions of the $X_i$ -- are computed and recorded. Our implementation supports saving and loading configurations from disk, allowing for checkpointing and reuse of thermalized states. 
To reduce autocorrelation and disk usage, we do not measure observables at every Monte Carlo step. Instead, measurements are taken every $\texttt{GAP}$ steps. In most of our simulations, we set \texttt{GAP} = 2.
The code monitors the acceptance rate and issues a warning if it falls below $50\%$, which may signal the need to adjust the integration step size $dt$ or the number of leapfrog steps.

\bibliographystyle{JHEP}
\bibliography{hoppe}
\end{document}